\documentclass[a4paper]{article}
\usepackage[T1]{fontenc}
\usepackage[english]{babel}
\usepackage{float}
\usepackage{bm}
\usepackage{amsmath}
\usepackage{amsthm}
\usepackage{graphicx}
\usepackage[]{color}

\makeatletter

\DeclareTextSymbolDefault{\textquotedbl}{T1}

\numberwithin{equation}{section}
\numberwithin{figure}{section}

\usepackage[all]{xy}
\usepackage{graphicx}

\makeatother

\begin{document}
\begin{flushleft} {\bf OUJ-FTC-8}\\ {\bf OCHA-PP-373}\\ \end{flushleft}

\begin{center} 

{\LARGE{}Fluctuating Non-linear Non-equilibrium System in Terms of
Nambu Thermodynamics}{\LARGE\par}

\vspace{32pt} 

So Katagiri\textsuperscript{{*}}\footnote{So.Katagiri@gmail.com},
Yoshiki Matsuoka$^{*}$, and Akio Sugamoto$^{\dagger}$
\begin{center}
\textit{$^{*}$Nature and Environment, Faculty of Liberal Arts, The
Open University of Japan, Chiba 261-8586, Japan}
\par\end{center}

\begin{center}
\textit{$^{\dagger}$Department of Physics, Graduate School of Humanities
and Sciences, Ochanomizu University, 2-1-1 Otsuka, Bunkyo-ku, Tokyo
112-8610, Japan }\\
\par\end{center}

\vspace{10pt}

\end{center}

\begin{abstract}
  It is shown that the structure of non-equilibrium thermodynamic systems
  far from equilibrium, can be captured in terms of a generalized ``Nambu
  dynamics'', in the presence of fluctuation effects
  in non-equilibrium thermodynamics. 
  \color{black}
  Triangular reactions are examined in detail, and it is shown that Nambu brackets can be used to describe them even when they are far from equilibrium, such as with cycles.
  Time evolution of the non-equilibrium state using the Hamiltonian and entropy is analyzed and it is shown that the entropy evolution is periodic with the negative contribution caused by the Hamiltonian suppressing the increase caused by entropy.
  \color{black}
  As concrete examples, chemical
  reaction systems with time oscillation, such as the Belousov-Zhabotinsky
  reaction (BZ reaction), \color{black}Hindmarsh-Rose (H-R) model\color{black}, are examined.
\end{abstract}

\section{Introduction }

To the non-equilibrium thermodynamic systems far from equilibrium,
the variational principle for entropy is not always applicable. The
construction of non-equilibrium thermodynamics, especially of non-linear
non-equilibrium thermodynamics requires going beyond the concept of
entropy \cite{key-3}\cite{key-4}; this is particularly evident in
systems such as chemical reactions. (The relationship with the non-linear
Kubo formula is summarized in \cite{key-5}.)

The Lotka-Volterra (LV) system of predator-prey relationship is one
such chemical reaction. The system can not be described by Onsager\textquoteright s
non-equilibrium thermodynamics. It is because Onsager\textquoteright s
thermodynamics takes the entropy as a potential, and requires ``the
minimum dissipative energy law'' to hold near the normal equilibrium
point\footnote{The same issues are discussed in \cite{key-65}.}.

According to \cite{key-10}, such Lotka-Volterra systems can be described
using the Nambu bracket \cite{key-9}. Here, we will refer to the non-equilibrium
thermodynamics described by such Nambu brackets as \textquotedblleft Nambu
non-equilibrium thermodynamics\textquotedblright .

In this paper, we will elucidate the structure of non-equilibrium
thermodynamic systems far from equilibria, by considering them as
`Nambu non-equilibrium thermodynamics\textquoteright 

\color{black}
\color{black} Nambu non-equilibrium thermodynamics consists of two ingredients; 1) the Nambu dynamics with multiple Hamiltonians $\{H_i; i=1, 2, \cdots\}$ and 2) the introduction of dissipation by a thermodynamic entropy $S$.  The first ingredient describes the flow of physical quantities following an incompressible fluid, while the second ingredient gives the non-equilibrium change of entropy within a compressible fluid. Therefore, this theory can be applied to the system either isolated, closed, or open.\color{black}\footnote{  
 For the closed system, we forbid the inflow and outflow of materials,  by imposing $\partial^{(H)}M_a=0$ ($M_a$: particle number of material $a$). For the isolated system, we further forbid the inflow and outflow of heat by $\partial^{(H)}S=0$.  For the open system no restrictions are imposed.
For usage of these symbols $\partial^{(H)}$, see Section 2.5.
\color{black}
} 

\color{black}The other advantage of the Nambu non-equilibrium thermodynamics is its capacity to handle non-linear systems and to describe phenomena such as oscillations (with cycles and spikes) through the interplay between multi-Hamiltonians and an entropy, which were previously inaccessible to conventional theories. \color{black}

We further construct a `fluctuating Nambu non-equilibrium thermodynamics\textquoteright{}
by adding to it the effect of fluctuations through ``quantization of
the Nambu brackets''. This method of quantization of the Nambu bracket
is applicable also to hydrodynamics and string (membrane) theory\cite{key-6}\cite{key-7},
since these theories are described by using the Nambu brackets. 

\color{black}
In the past several decades, numerous attempts have been made to extend Onsager's near-equilibrium thermodynamics to far-from-equilibrium systems.
Among these, the Glansdorff-Prigogine theory is a well-established framework \cite{Prigogine}. However, this theory is not so powerful to capture oscillatory phenomena or to provide a general principle for determining far-from-equilibrium systems. Comparison of \cite{Prigogine} and other attempts with our paper, is 
given in the last section 6.\color{black}

The paper is structured as follows. The Nambu bracket is first explained
in Section 2. Next in Section 3, derivation of the general form for Nambu non-equilibrium thermodynamics 
is discussed. In Section 4, approaches to construct the Hamiltonians and entropy are described.
In the following Section 5, it is given how the BZ reaction and H-R model
is described by the Nambu bracket and the Nambu non-equilibrium thermodynamics,
in which an entropy term is introduced in addition to two Hamiltonians.
\color{black} In Section 6, we discuss how the characteristic cycles and spikes in Nambu non-equilibrium thermodynamics are realized using two toy models.
In Section 7, we discuss the effect of fluctuations on Nambu non-equilibrium
thermodynamics, and
numerically investigate how the effect of fluctuations affects the
time oscillation-like behavior.
\color{black} Finally, in Section 6, the results
are summarized and discussed.

\color{black}In addition, Appendix A discusses entropy in LCR circuits.
Appendix B examines spikes in more detail, and Appendix C discusses the Zwanzig model.
Appendix D discusses \color{black} the specific example of moving toward the thermal equilibrium, and Appendix E discusses harmonic oscillators.
\color{black}

\color{black} Before ending this section, we will repeat our motivation to write this paper: \color{black}This paper intends to understand the periodic motions (cycles) in various models far from the equilibrium. We separate the motion into the ordinary part of periodic motion and the dissipation part. The former part is supported by the Nambu dynamics with multiple conserved quantities (Hamiltonians), and the other is supported by the entropy effects.  The inclusion of the entropy effects lifts the Nambu dynamics to the Nambu non-equilibrium thermodynamics.

\section{Nambu non-equilibrium thermodynamics}

\subsection{Nambu bracket}

In this subsection we discuss the Nambu bracket \cite{key-9}. \color{black}
In the Hamiltonian dynamical system, phase space is described by two variables, position and momentum, and its equation of motion is written as

\begin{equation}
  \dot{x}^{i}=-\{H,x^{i}\}_{PB},
\end{equation}
where $H$ is the Hamiltonian, and $\{H,x^{i}\}_{PB}$ is the Poisson bracket:
\begin{equation}
  \{A_{1},A_{2}\}_{PB}\equiv\epsilon^{i_{1}i_{2}}\frac{\partial A_{1}}{\partial x^{i_{1}}}\frac{\partial A_{2}}{\partial x^{i_{2}}}.
\end{equation}
A feature of Hamiltonian dynamics is that the volume element in phase space is invariant with time variation, known as Liouville's theorem.
Since one can derive principle of equal a priori probabilities by assuming ergodic properties if Liouville's theorem is satisfied, Nambu discussed in 1973 the existence of a more general dynamics that satisfies the more general Liouville's theorem which can reproduce the microscopic theory behind statistical mechanics.
\color{black}

Nambu bracket is a multi-variable extension of the Poisson bracket in the usual Hamiltonian
mechanics, and can be written as a multi-variable Jacobian:

\begin{equation}
\{A_{1},A_{2},\dots,A_{n}\}_{NB}\equiv\epsilon^{i_{1}i_{2}\dots i_{n}}\frac{\partial A_{1}}{\partial x^{i_{1}}}\frac{\partial A_{2}}{\partial x^{i_{2}}}\dots\frac{\partial A_{n}}{\partial x^{i_{n}}}.
\end{equation}
If $n=2$, the Nambu brackets give the usual Poisson brackets.

By using the Nambu bracket, the canonical equation in Nambu mechanics
is extended to

\begin{equation}
\dot{x}^{i}=-\{H_{1},\dots,H_{n-1},x^{i}\}_{NB}= -\epsilon^{i_{1}\dots i_{n-1}i}\frac{\partial H_{1}}{\partial x^{i_{1}}}\dots\frac{\partial H_{n-1}}{\partial x^{i_{n-1}}}.
\end{equation}

\color{black}
In addition to being able to describe dynamical systems that cannot be described by ordinary Hamiltonian mechanics, Nambu brackets have the advantage of being able to explicitly describe conserved quantities, in the form of multiple Hamiltonians for ordinary Hamiltonian dynamical systems.

As a specific example, in the rotational motion of a rigid body, $L_{i}$ is the angular momentum, $I_{i}$ is the moment of inertia, and the total kinetic energy $G$,
\begin{equation}
  G=\frac{1}{2}\left(\frac{L_{x}^{2}}{I_{x}}+\frac{L_{y}^{2}}{I_{y}}+\frac{L_{z}^{2}}{I_{z}}\right)
\end{equation}
and the square of the angular momentum $H$, 
\begin{equation}
  H=\frac{1}{2}\left(L_{x}^{2}+L_{y}^{2}+L_{z}^{2}\right)
\end{equation}
are ``two'' conserved quantities in the time evolution, \color{black}which can be used to write the Euler equations for the rigid body as 
\begin{equation}
  \dot{L}_{i}=\{L_{i},H,G\}_{NB}.
\end{equation}

In the second example, the Nambu bracket representation of hydrodynamics is also known.
Introducing Hamiltonians $H_1$ and $H_2$ like 
\begin{equation}
  \dot{x}^{i}=-\{x^{i},H_{1},H_{2}\}_{NB}
\end{equation}
for the fluid elements, the advection terms can be expressed in Nambu brackets as in 
\begin{equation}
  \{\{x^{i},H_{1}, H_{2}\},H_{1},H_{2}\}_{NB}=\{v^{i},H_{1},H_{2}\}_{NB}=\frac{\partial v^{i}}{\partial x^{j}}\{x^{j},H_{1},H_{2}\}_{NB}=v^{j}\frac{\partial v^{i}}{\partial x^{j}},  
\end{equation}
where Hamiltonians $H_1, H_2$ are equal to stream functions in hydrodynamics $\psi_1, \psi_2$\footnote{In the $D$-dimensional hydrodynamics, there are $D-1$ stream functions, $\psi_1,\dots, \psi_{D-1}$, which can be considered as ``multiple Hamiltonians'', $H_1,\dots,H_{D-1}$, respectively.}.
Therefore, the Navier-Stokes equation 
\begin{equation}
  \rho\left(\frac{\partial}{\partial t}v^{i}+v^{j}\frac{\partial v^{i}}{\partial x^{j}}\right)=-\frac{\partial}{\partial x^{i}}p+\eta\frac{\partial^{2}}{\partial x^{2}}v^{i}
\end{equation}
can be written as 
\begin{align}
  -&\rho\left(\{x^{i},\dot{H_{1}},H_{2}\}_{NB}+\{x^{i},H_{1},\dot{H}_{2}\}_{NB}+\{\{x^{i},H_{1},H_{2}\}_{NB},H_{1},H_{2}\}_{NB}\right) \\
  &=\epsilon_{ijk}\{p,x^{j},x^{k}\}_{NB}-\eta\frac{\partial^{2}}{\partial x^{2}}\{x^{i},H_{1},H_{2}\}_{NB}
\end{align}
using Nambu brackets.

\color{black}

\color{black}
\subsection{Onsager's non-equilibrium thermodynamics}

In this subsection, we discuss the Onsager\textquoteright s thermodynamics\cite{key-3}\cite{key-gaugefixing}.
Thermodynamic variables evolve in time by gradients of entropy, 

\begin{equation}
\dot{x}^{i}=L^{ij}\frac{\partial S}{\partial x^{j}}.
\end{equation}
where $x^i$ is a thermodynamic variable, e.g., $x^i=E,V,N,\dots$.

The so-called ``stochastic entropy'' $S_{\mathrm{OMH}}$, Shannon entropy of probability density according to Onsager, Machlup and Hashitsume (Onsager-Machlup-Hashitsume action) \cite{key-100}\cite{key-101}\cite{key-102}\cite{key-103},
can be written as
\begin{equation}
  \dot{S}_{\mathrm{OMH}} = \Phi - \dot{x}^i A_i + \Psi = \frac{1}{2}R_{ik} (\dot{x}^i - L^{ij}A_j)(\dot{x}^k -L^{kl}A_l) \geq 0,
\end{equation}
where $\Phi$ is the first-class dissipation function and $\Psi$ is the second-class dissipation,
\begin{equation}
  \Phi = \frac{1}{2}R_{ij} \dot{x}^i \dot{x}^j, \Psi = \frac{1}{2} L^{ij} A_i A_j
\end{equation}
and $A_i$ is $A_i = \frac{\partial S}{\partial x^i}$ and  $R_{ij}$ is the inverse
matrix of $L^{ij}$.\footnote{
  \color{black}
  The discussion in Section 2.2 is limited to isolated systems. However,
  the argument itself can be applied to both closed and open systems, at least when the thermal fluctuations added to the system are Gaussian. In that case, we can obtain a path integral representation consistent with the boundary condition of each system.
\color{black}
}

Since $S_\mathrm{OMH}$ is Shannon entropy itself, the probability distribution is described as
\begin{equation}
  P(x,t_2|x',t_1) = \int Dx \; e^{-\frac{1}{k_B}\int^{t_2}_{t_1}\dot{S}_{\mathrm{OMH}}dt}=\int Dx \;  e^{-\frac{1}{k_B}S_{\mathrm{OMH}}(x,t_2|x',t_1)}
\end{equation}
by path integrals.  Here,  $S_{\mathrm{OMH}}(x,t_2|x',t_1)= \int_{t_1, x(t_1)=x'}^{t_2, x(t_2)=x} dt \; \dot{S}_{\mathrm{OMH}}$. 

Therefore, the probability distribution is most effective when the generation rate of stochastic entropy $\dot{S}_{\mathrm{OMH}}$ is at its minimum, at which point Onsager's canonical equation 
\begin{equation}
  \dot{x}^i = L^{ij} A_j
\end{equation}
becomes valid (Principle of minimum dissipation of entropy).
In fact, other paths can be considered, which gives fluctuation to the path.

Under $\dot{S}_{OMH}=0$, the entropy $S$ satisfies
\begin{equation}
  \dot{S} = \dot{x^i}A_i= \Phi + \Psi \geq 0.
\end{equation}
Therefore, the entropy increases continuously in time, when $R_{ij}$ is a positive semi-definite matrix.

The second-order derivative of entropy in time is

\begin{equation}
\begin{aligned}\ddot{S} & =\left(2\frac{\partial^{2}S}{\partial x^{i}\partial x^{j}}+R_{il}\dot{L}^{lk}R_{kj}\right)\dot{x}^{i}\dot{x}^{j}\\
 & \equiv\left(2g_{R,ij}+R_{il}\dot{L}^{lk}R_{kj}\right)\dot{x}^{i}\dot{x}^{j}\\
 & \equiv g_{ij}\dot{x}^{i}\dot{x}^{j}
\end{aligned}
\end{equation}
where $g_{R,ij}$ is the Ruppeiner metric.

Thus, the acceleration of entropy is determined by $g_{ij}$, and it
is reasonable to assume that fluctuations satisfy

\begin{equation}
g_{ij}\dot{x}^{i}\dot{x}^{j}<0
\end{equation}
near equilibrium points.

Thus, the magnitude of the entropy increase continues to decrease,
reaching an equilibrium state, where the entropy increase ceases.

A significant assumption in Onsager's non-equilibrium thermodynamics is the assumption that the time evolution $\dot{x}^i$ of the thermodynamic variable is linearly proportional to the thermodynamic force $A_i$.
Onsager argued that by assuming reversibility in the microscopic theory behind non-equilibrium thermodynamics, the transport coefficients could exhibit symmetries such as $L^{ij}=L^{ji}$. As a typical example of this, he discussed the triangular reaction in the near-equilibrium case.
Onsager's non-equilibrium thermodynamics is based on the assumption of the ``dynamic reversibility principle''.   Since the cyclical reactions in the triangular reaction are not reversible in time, the Onsager's theory works under the condition that the cyclical reactions do not occur.  \color{black}As we see in the following, such an assumption only covers near-equilibrium situations.   On the other hand, Nambu non-equilibrium thermodynamics can treat a cycle naturally, so that it can be applied to phenomena far from equilibrium.\color{black}

\subsection{Triangular reaction}
Triangular reactions were first discussed by Onsager when he constructed non-equilibrium thermodynamics\cite{key-3}. In this discussion, we can understand that Onsager's assumptions and his theory are valid only in near-equilibrium situations.

A triangular reaction is a chemical reaction, in which three chemical species $X_1,X_2$ and $X_3$ make transition to each other.
\begin{equation}
\xymatrix{X_{1}\ar[r]\ar[d] & X_{2}\ar[ld]\ar[l]\\
X_{3}\ar[u]\ar[ru]
}
\end{equation}
 This can generally be extended to $n$ species.
Notice that the reaction contains a right-handed cycle 
\begin{equation}
\xymatrix{X_{1}\ar[r] & X_{2}\ar[ld]\\
X_{3}\ar[u]
}
\end{equation}
and a left-handed cycle.
\begin{equation}
\xymatrix{X_{1}\ar[d] & X_{2}\ar[l]\\
X_{3}\ar[ru]
}        
\end{equation}
These cycles are not necesarilly reversible with respect to time reversal in individual cycles.
If we write the rate constant for the reaction as $k_{ij}$, the triangular reaction can be decomposed as 
\begin{align}
X_{1}\overset{k_{12}}{\underset{k_{21}}{\rightleftharpoons}}X_{2}, \\
X_{2}\overset{k_{23}}{\underset{k_{32}}{\rightleftharpoons}}X_{3}, \\
X_{3}\overset{k_{31}}{\underset{k_{13}}{\rightleftharpoons}}X_{1}.
\end{align}

Since the rate of reaction is proportional to the concentration, the time evolution of the thermodynamic variables is given by following equations.
\begin{align}
    \frac{dx_{1}}{dt}=-\left(k_{12}+k_{13}\right)x_{1}+k_{21}x_{2}+k_{31}x_{3}=-k_{12}x_{1}\left(1-\frac{k_{21}}{k_{12}}\frac{x_{2}}{x_{1}}\right)+k_{31}x_{3}\left(1-\frac{k_{13}}{k_{31}}\frac{x_{1}}{x_{3}}\right) \\
    \frac{dx_{2}}{dt}=-\left(k_{23}+k_{21}\right)x_{2}+k_{32}x_{3}+k_{12}x_{1}=-k_{23}x_{2}\left(1-\frac{k_{32}}{k_{23}}\frac{x_{3}}{x_{2}}\right)+k_{12}x_{1}\left(1-\frac{k_{21}}{k_{12}}\frac{x_{2}}{x_{1}}\right) \\    
    \frac{dx_{3}}{dt}=-\left(k_{31}+k_{32}\right)x_{3}+k_{13}x_{1}+k_{23}x_{2}=-k_{31}x_{3}\left(1-\frac{k_{13}}{k_{31}}\frac{x_{1}}{x_{3}}\right)+k_{23}x_{2}\left(1-\frac{k_{32}}{k_{23}}\frac{x_{3}}{x_{2}}\right)
\end{align}

Now consider an ideal gas, and from the Gibbs free energy $G(T,p,N)$, the chemical potential of the ideal gas has the relation 
\begin{equation}
\beta d\mu=d\log p.
\end{equation}
From the equation of state of the ideal gas $p V = N \beta^{-1}$, we obtain
\begin{equation}
\frac{x_i}{x_j} = e^{\beta(A_{ij}-A_{(0)ij})}
\end{equation}

\color{black}
where $A_{ij}$ is defined as $A_{ij} = \mu_i - \mu_j$ and $A_{(0)ij}$ is the equilibrium value of $A_{ij}$ and $x_i = N_i/V$.
Here $x_i=x_j$ holds when the system is in equilibrium, i.e. $A_{ij}=A_{(0)ij}$.
\color{black}

Substituting these, the time evolution of the thermodynamic variables is given by

\begin{align}
\frac{dx_{1}}{dt}=-k_{12}\left(1-\frac{k_{21}}{k_{12}}e^{\beta(A_{12}-A_{(0)21})}\right)x_{1}+k_{31}\left(1-\frac{k_{13}}{k_{31}}e^{\beta(A_{31}-A_{(0)31})}\right)x_{3}, \\
\frac{dx_{2}}{dt}=-k_{23}\left(1-\frac{k_{32}}{k_{23}}e^{\beta(A_{23}-A_{(0)23})}\right)x_{2}+k_{12}\left(1-\frac{k_{21}}{k_{12}}e^{\beta(A_{12}-A_{(0)12})}\right)x_{1}, \\
\frac{dx_{3}}{dt}=-k_{31}\left(1-\frac{k_{13}}{k_{31}}e^{\beta(A_{31}-A_{(0)31})}\right)x_{3}+k_{23}\left(1-\frac{k_{32}}{k_{23}}e^{\beta(A_{23}-A_{(0)23})}\right)x_{2}. 
\end{align}

It is important to note here that the time evolution equation obtained is not at all linear in the thermodynamic force $A_{ij}$. In other words, general triangular reactions cannot be discussed at all in Onsager's linear non-equilibrium thermodynamics.
\color{black}
It is applicable only when the affinity is small in the first approximation in $\beta$ of the exponential function.
Under this approximation, the time evolution is linear about the chemical potential plus cycle part.  

\begin{align}
\frac{dx_{1}}{dt}=-k_{12} x_1 + k_{21} x_1 + k_{21} \beta (A_{12} - A_{(0)12}) x_1 \\
\frac{dx_{2}}{dt}=-k_{23} x_2 + k_{32} x_2 + k_{32} \beta (A_{23} - A_{(0)23}) x_2 \\
\frac{dx_{3}}{dt}=-k_{31} x_3 + k_{13} x_3 + k_{13} \beta (A_{31} - A_{(0)31}) x_3 \\
\end{align}

The cycle part is usually eliminated by assuming a detailed balance principle near equilibrium.

\begin{align}
-k_{12} x_1 + k_{21} x_2 = 0 \\
-k_{23} x_2 + k_{32} x_3  = 0\\
-k_{31} x_3 + k_{13} x_1  = 0\\
\end{align}
\color{black}
Therefore, when the detail balance is valid,   the time evolution can be described by Onsager's non-equilibrium thermodynamics.

\begin{align}
\frac{dx_{1}}{dt}=-\beta\left(k_{12}x_{1}+k_{13}x_{3}\right)\left(\mu_{1}-\mu_{1(0)}\right)+\beta k_{21}x_{1}\left(\mu_{2}-\mu_{2(0)}\right)+\beta k_{13}x_{3}(\mu_{3}-\mu_{3(0)}) \\
\frac{dx_{2}}{dt}=-\beta\left(k_{23}x_{2}+k_{21}x_{1}\right)\left(\mu_{2}-\mu_{2(0)}\right)+\beta k_{32}x_{2}\left(\mu_{3}-\mu_{3(0)}\right)+\beta k_{21}x_{1}(\mu_{1}-\mu_{1(0)}) \\
\frac{dx_{3}}{dt}=-\beta\left(k_{31}x_{3}+k_{32}x_{2}\right)\left(\mu_{3}-\mu_{3(0)}\right)+\beta k_{13}x_{3}\left(\mu_{1}-\mu_{1(0)}\right)+\beta k_{32}x_{2}(\mu_{2}-\mu_{2(0)})
\end{align}
where $\mu_{i(0)}$ is the equilibrium value of the chemical potential $\mu_{i}$.
 
In this case, the transport coefficient is obtained as

\begin{align}
L^{11}=\beta\left(k_{12}x_{1}+k_{13}x_{3}\right), \\
L^{22}=\beta\left(k_{23}x_{2}+k_{21}x_{1}\right), \\
L^{33}=\beta\left(k_{31}x_{3}+k_{32}x_{2}\right), \\
L^{12}=-\beta k_{21}x_{1}=L^{21}, \\
L^{23}=-\beta k_{32}x_{2}=L^{32}, \\
L^{31}=-\beta k_{13}x_{3}=L^{13}.
\end{align}

In this way, the triangular reaction is described by Onsager's non-equilibrium thermodynamics, when it is close to equilibrium and without cycles.
However, Onsager's non-equilibrium thermodynamics cannot describe cases that are far from equilibrium, such as those involving cycles.

\color{black}
\subsection{ Nambu mechanics as a higher-order approximation
beyond Onsager non-equilibrium thermodynamics}

In the previous subsection we derived the transport coefficients for
Onsager's nonequilibrium thermodynamics by taking a first-order approximation
with small affinity. Also, the 0th approximation was dropped by hand
according to \textquotedbl detailed balance principle\textquotedbl . 

If these assumptions and approximations are not adopted, the expansion
of affinity yields the following expression.

\begin{equation}
\begin{aligned}\frac{dx_{1}}{dt} & =-k_{12}x_{1}+k_{31}x_{3}\\
 & +\sum_{n=0}^{\infty}\sum_{k=0}^{n}\frac{1}{k!(n-k)!}\beta^{n}\left(-1\right)^{n-k}\left(k_{21}x_{1}\Delta\mu_{1}^{k}\Delta\mu_{2}^{n-k}-k_{13}x_{3}\Delta\mu_{3}^{k}\Delta\mu_{1}^{n-k}\right)
\end{aligned}
\end{equation}

\begin{equation}
\begin{aligned}\frac{dx_{2}}{dt} & =-k_{23}x_{2}+k_{12}x_{1}\\
 & +\sum_{n=0}^{\infty}\sum_{k=0}^{n}\frac{1}{k!(n-k)!}\beta^{n}\left(-1\right)^{n-k}\left(k_{32}x_{2}\Delta\mu_{2}^{k}\Delta\mu_{3}^{n-k}-k_{21}x_{1}\Delta\mu_{1}^{k}\Delta\mu_{2}^{n-k}\right)
\end{aligned}
\end{equation}

\begin{equation}
\begin{aligned}\frac{dx_{3}}{dt} & =-k_{31}x_{3}+k_{23}x_{2}\\
 & +\sum_{n=0}^{\infty}\sum_{k=0}^{n}\frac{1}{k!(n-k)!}\beta^{n}\left(-1\right)^{n-k}\left(k_{13}x_{3}\Delta\mu_{3}^{k}\Delta\mu_{1}^{n-k}-k_{32}x_{2}\Delta\mu_{2}^{k}\Delta\mu_{3}^{n-k}\right)
\end{aligned}
\end{equation}
where $\Delta\mu_{i}\equiv\mu_{i}-\mu_{(0)i}.$

Let us define the coefficients of each term of the chemical potential
as follows;

\begin{equation}
\frac{dx_{i}}{dt}=\sum_{n=0}^{\infty}\sum_{i_{1},\dots i_{n}=1}^{3}L_{i,i_{1}i_{2}i_{3}\dots i_{n}}\Delta\mu_{i_{1}}\dots\Delta\mu_{i_{n}}.
\end{equation}

From the definition, $L_{i,i_{1}i_{2}i_{3}\dots i_{n}}$ is symmetric
with respect to $i_{1},\dots,i_{n}.$

In fact, if we expand from lower orders;

\begin{equation}
\frac{dx_{i}}{dt}=v_{i}^{(0)}+v_{i}^{(1)}+v_{i}^{(2)}+\dots,
\end{equation}
we obtain

\begin{equation}
v_{i}^{(0)}=L_{i},
\end{equation}

\begin{equation}
v_{i}^{(1)}=\sum_{i_{1}=1}^{3}L_{i,i_{1}}\Delta\mu_{i_{1}},
\end{equation}

\begin{equation}
v_{i}^{(2)}=\sum_{i_{1},i_{2}=1}^{3}L_{i,i_{1}i_{2}}\Delta\mu_{i_{1}}\Delta\mu_{i_{2}}
\end{equation}
where $L_{i}$ are 

\begin{equation}
L_{1}=-(k_{12}-k_{21})x_{1}+\left(k_{31}-k_{13}\right)x_{3},
\end{equation}

\begin{equation}
L_{2}=-(k_{23}-k_{32})x_{2}+\left(k_{12}-k_{21}\right)x_{1},
\end{equation}

\begin{equation}
L_{3}=-(k_{31}-k_{13})x_{3}+\left(k_{23}-k_{32}\right)x_{2}
\end{equation}
and $L_{i,j}$ are the same as $L_{ij}$ (2.38)-(2.43).

We focus on second-order approximations.

$L_{i,i_{1}i_{2}}$ are 

\begin{equation}
L_{1,11}=\frac{1}{2}\beta^{2}\left(k_{21}x_{1}-k_{13}x_{3}\right),\ L_{1,22}=\frac{1}{2}\beta^{2}k_{21}x_{1},L_{1,33}=-\frac{1}{2}\beta^{2}k_{13}x_{3}
\end{equation}

\begin{equation}
L_{1,12}=-\frac{1}{2}\beta^{2}k_{21}x_{1},\ L_{1,13}=\frac{1}{2}\beta^{2}k_{13}x_{3}
\end{equation}

\begin{equation}
L_{2,22}=\frac{1}{2}\beta^{2}\left(k_{32}x_{2}-k_{21}x_{1}\right),\ L_{2,11}=-\frac{1}{2}\beta^{2}k_{21}x_{1},L_{2,33}=\frac{1}{2}\beta^{2}k_{32}x_{2},
\end{equation}

\begin{equation}
L_{2,23}=-\frac{1}{2}\beta^{2}k_{32}x_{2},\ L_{2,21}=\frac{1}{2}\beta^{2}k_{21}x_{1},
\end{equation}

\begin{equation}
L_{3,33}=\frac{1}{2}\beta^{2}\left(k_{13}x_{3}+k_{32}x_{2}\right),\ L_{3,11}=\frac{1}{2}\beta^{2}k_{13}x_{3},L_{3,22}=-\frac{1}{2}\beta^{2}k_{32}x_{2},
\end{equation}

\begin{equation}
L_{3,31}=-\frac{1}{2}\beta^{2}k_{13}x_{3},\ L_{3,32}=\frac{1}{2}\beta^{2}k_{32}x_{2}.
\end{equation}

From these we see that $L_{i,i_{1}i_{2}}$ has the following symmetry
with respect to the index;

\begin{equation}
L_{i,i_{1}i_{2}}=L_{i_{1},ii_{2}}=L_{i_{2},ii_{1}}.
\end{equation}

In addition to this symmetry, it is noteworthy that
$L_{i,i_{1}i_{2}}$ has the following antisymmetry.

\begin{equation}
L_{i,jj}=-L_{j,ii}
\end{equation}
where no sum is taken for the subscripts that appear twice here.

Therefore, we introduce the following new symmetric tensor;

\begin{equation}
\tilde{A}_{(ij)}\equiv\frac{1}{2}(\Delta\mu_{i}\Delta\mu_{j}+\Delta\mu_{j}\Delta\mu_{i}).
\end{equation}

We then introduce the antisymmetric tensor as follows

\begin{equation}
B_{i,j}\equiv L_{i,jj}.
\end{equation}

Thus $v_{i}^{(2)}$ is written as 

\begin{equation}
v_{i}^{(2)}=\sum_{j=1}^{3}B_{i,j}\tilde{A}_{(jj)}+\sum_{j=1}^{3}L_{i,ij}\tilde{A}_{(ij)}.
\end{equation}

They can be described in the form of Nambu non-equilibrium thermodynamics
using $H_{1}^{(2)},H_{2}^{(2)}$ and $S^{(2)}$ as follows;

\begin{equation}
H_{1}^{(2)}\equiv\sum_{i=1}^{3}\tilde{A}_{(ii)}x^{i},
\end{equation}

\begin{equation}
H_{2}^{(2)}= \beta^2(k_{21} x^1 x^3 + k_{32} x^2 x^1 + k_{13} x^3 x^2)
\end{equation}

\begin{equation}
S^{(2)}\equiv\sum_{i,j=1}^{3}\tilde{A}_{(ij)}x^{i}x^{j},
\end{equation}

\begin{equation}
v_{i}^{(2)}=\sum_{j,k=1}^{3}\epsilon^{ijk}\frac{\partial H_{1}^{(2)}}{\partial x^{j}}\frac{\partial H_{2}^{(2)}}{\partial x^{j}}+\sum_{k=1}^{3}L_{i,ik}\frac{\partial}{\partial x^{i}}\left(\frac{\partial S^{(2)}}{\partial x^{k}}\right)
\end{equation}

where $H^{(2)}_2$ is determinded by $B_{ij}$;
\begin{equation}
  \epsilon^{ijk} \frac{\partial H^{(2)}_2}{\partial x^k} \equiv B_{ij}.
\end{equation}

Thus, it can be seen that the terms of Nambu mechanics emerge when
we consider the second order terms of the approximation.

The zeroth order terms are similarly described by H and S, and are
as follows,

\begin{equation}
H_{1}^{(0)}\equiv\frac{1}{2}(k_{12}-k_{21})x_{3}+\frac{1}{2}(k_{31}-k_{13})x_{2}+\frac{1}{2}(k_{23}-k_{32})x_{1}
\end{equation}

\begin{equation}
H_{2}^{(0)}\equiv\frac{1}{2}\left(x_{1}^{2}+x_{2}^{2}+x_{3}^{2}\right)
\end{equation}

\begin{equation}
S^{(0)}\equiv-\frac{1}{2}\left((k_{12}-k_{21})x_{1}(x_{1}+x_{2})+(k_{23}-k_{32})x_{2}(x_{2}+x_{3})+(k_{31}-k_{13})x_{3}(x_{3}+x_{1})\right)
\end{equation}

\begin{equation}
v_{i}^{(0)}=\sum_{j,k=1}^{3}\epsilon^{ijk}\frac{\partial H_{1}^{(0)}}{\partial x^{j}}\frac{\partial H_{2}^{(0)}}{\partial x^{j}}+\frac{\partial S^{(0)}}{\partial x^{i}}
\end{equation}

From these, the triangular reaction is described by the second-order
approximation as follows;

\begin{equation}
\frac{dx^{i}}{dt}=\epsilon^{ijk}\frac{\partial H_{1}^{(0)}}{\partial x^{j}}\frac{\partial H_{2}^{(0)}}{\partial x^{k}}+\epsilon^{ijk}\frac{\partial H_{1}^{(2)}}{\partial x^{j}}\frac{\partial H_{2}^{(2)}}{\partial x^{k}}+\frac{\partial S^{(0)}}{\partial x^{i}}+L_{ij}\frac{\partial S^{(1)}}{\partial x^{j}}+L_{i,ij}\frac{\partial}{\partial x^{i}}\frac{\partial S^{(2)}}{\partial x^{j}}+\dots
\end{equation}
where $S^{(1)}$ is 
\begin{equation}
\frac{\partial S^{(1)}}{\partial x^{i}}\equiv\Delta\mu_{i}
\end{equation}
in the usual Onsager theory.

Thus, Nambu non-equilibrium thermodynamics naturally appears as a
correction term when an approximation is given without necessarily
assuming detailed balance.
\color{black}

\subsection{Nambu non-equilibrium thermodynamics}

\color{black}
To introduce the Nambu non-equilibrium thermodynamics, it is natural to start from a set of autonomous type differential equations for
$N$ thermodynamic variables, $\left(x^i \right)=\left(x^{1},x^{2},\cdot\cdot\cdot, x^{N}\right)$. \color{black}
 The thermodynamic variables can be
anything, such as $(E,V,N_{1},\cdot\cdot\cdot,N_{s})$ or $(T,p,\mu_{1},\cdot\cdot\cdot,\mu_{s})$
for chemical reactions, and others.\footnote{The variables can accommodate gauge potential $A_{\mu}(x)$ or electromagnetic
fields$(E,B)$ for electromagnetic reactions, potential of bio-membrane
and concentration of ions for the signal propagation problem of the
nerve, and so on.}

The differential equation of our autonomous system can be written as

\begin{equation}
\frac{dx^{i}(t)}{dt}=v^{i}(\bm{x})\ (i=1,2,\dots,N).
\end{equation}

Then, a velocity field $v={v^{i}(\bm{x})}$ is given, it determines
a ``fluid flow'' in the $N$-dimensional space $\{x\}$, called here the
``thermodynamic space''.

As is well known in the vector analysis or in the hydrodynamics, any
velocity field $\bm{v}(\bm{x})$ can be expressed as a sum of divergence-free part $\bm{v}^{(1)}(\bm{x})$ and a rotation-free part $\bm{v}^{(2)}(\bm{x})$.  This is called the Helmholtz theorem in physics, or Hodge theorem in mathematics.

For $N=3$ we can find stream functions $\psi_i(x)$ and a velocity function $\phi(x)$ such that 
\begin{equation}
v^{i}(\bm{x})=(\bm{v}^{(1)}(\bm{x})+\bm{v}^{(2)}(\bm{x}))^{i}=\sum_{j,k=1}^{3}\epsilon^{ijk}\partial_{j}\psi_{k}(\bm{x})+\delta^{ij}\partial_{j}\phi(\bm{x}).
\end{equation}

For a general dimension $N$, the role of stream functions is played by a two-form field $\psi^{ij}(\bm{x})$ with $\psi^{ij}=-\psi^{ji}$, which implies $\psi^{ij}=\epsilon^{ijk}\partial_{j}\psi_{k}$ for $N=3$.

The first term $\bm{v}^{(1)}(\bm{x})$ in the right-hand side satisfies the
incompressible condition $\bm{\nabla}\cdot\bm{v}^{(1)}=0$, while
the second term $\bm{v}^{(2)}(\bm{x})$ gives the compressible flow of
the fluid. 

We expect that the first term of the incompressible flow can be expressed, even for a general $N$, in terms of $N-1$ ``Hamiltonians'' $(H_{1},H_{2},\dots,H_{N-1})$ as 
\begin{equation}
\text{\ensuremath{\bm{v}}}^{(1)}(\bm{x})=\{\bm{x},H_{1}(\bm{x}),\dots,H_{N-1}(\bm{x})\}_{NB}.
\end{equation}

This expectation is true \underline{locally} (for a short time and in a small space region). See a proof of the ``local existence theorem'' in the footnote \footnote{A mathematical version of the Helmholtz theorem (the Hodge theorem) states $^\exists \psi^{\mu\nu}(x)$ (2-form field) s.t. $v^{(1)\mu}=\partial_{\nu}\psi^{\mu\nu}$, in the flat $x$-space. Next, the Darboux theorem\cite{Darboux} gives a canonical frame $\{y^i\}$, in which, by coordinate transformation, the 2-form field can be reduced to a constant skew symmetric one, $\psi^{'ij}=\sum_{k=1}^p \left( \delta^i_ {2k-1}\delta^j_{2k}-\delta^i_{2k}\delta^j_{2k-1} \right)$, where $^\exists p$ s.t. $2p \le N$. Then, the corresponding dual $(N-2)$-form reads
$$
\psi^*_{\mu_3 \cdots \mu_N}(x) =f(y)   \epsilon^{(0)}_{ij i_3 \cdots i_N}  \left( \frac{\partial y^{i_3}}{\partial x^{\mu_3}} \cdots \frac{\partial y^{i_N}}{\partial x^{\mu_N}} \right),
$$
where we have introduced $f(y)$ by relaxing the choice of $y$-frame. The epsilon symbol with $(0)$ means $\pm 1$ depending on even and odd permutations of the indices.
Then, we can identify Nambu Hamiltonians $(H_1, \cdots, H_{N-1})$ by $y^{i+1} = \ln H_{i}(x)$ and $f(y)= \prod_{i=1}^{N-1} H_i(x)$, which leads to 
$$
\psi^*_{\mu_3 \cdots \mu_N}(x)=  \sum_{i_1, \cdots, i_N \in (1, \cdots, N-1)}
 \epsilon^{(0)}_{i_1 \cdots i_{N-1}}    \left( H_{i_1} \frac{ \partial H_{i_2}}{\partial x^{\mu_3}}  \cdots \frac{\partial H_{i_{N-1}}}{\partial x^{\mu_N}}\right),
$$
Now, we arrive at the same expression $\dot{x}^{\mu}(t)$ in Nambu dynamics, at least locally:
$$
v^{(1)\mu}(x)=\frac{1}{(N-2)!} \epsilon_{(0)}^{\mu\nu\mu_3 \cdots \mu_N} \partial_{\nu} \psi^*_{\mu_3 \cdots \mu_N} =\{x^{\mu}, H_1, H_2, \cdots, H_{N-1}\}_{NB}.
$$ The determination of $H's$ is equivalent to the finding of the reduced $y$-frame, so that the local existence theorem works irrespective of whether the entropy $S(\bm{x})$ exists or not. ({\it q.e.d.})
} 

This local existence theorem is, however, frequently broken by the occurrence of chaotic and other singular behaviors of the solution. Therefore, the globalization of the local existence theorem is a difficult problem and remains still an unsolved problem.

Many counter-examples of the global non-existence of  $H's$ are known. A well-known example is the Kowalevskaya top, which is an in-compressible version of the autonomous system with 6 variables ($N=6$).  In this three-dimensional top moving under the gravity, any choice of parameters other than those chosen by Euler, Lagrange and Kowalevskaya, the number of ``Hamiltonians'' obtained is four, one less than the maximum $N-1=5$. 

In this situation, we are better to formulate the Nambu non-equilibrium thermodynamics under the following assumptions:

$\bullet$ \underline{Assumptions}: Assume the existence of $N-1$ ``Hamiltonians'' $(H_1, \cdots, H_{N-1})$, so that $H_i(x_1. \cdots, x_N)=$ const may determine a hyper-surface in the $N$-dimensional space, and that these hyper-surfaces are independent with each other, or $(N-1)$ tangential vectors $(\bm{\nabla}H_1, \cdots, \bm{\nabla}H_{N-1})$ are linearly independent.
We also assume $H_i(\bm{x})_{i=1-(N-1)}$ are the $C^{\infty}$-class or the real analytic functions in the $N$-dimensional space. 

The incompressibility implies the volume of the thermodyanamic
space is conserved, that is, this part gives the dynamics where
the Liouville theorem and its generalization hold. This is a merit of the Nambu dynamics. On the other hand,
in the compressibility case, the violation of the Liouville theorem occurs; the volume is expanded or contracted, giving the entropy effect. The degree of expansion and contraction is given by the Lyapunov exponent in mathematics, so that the entropy can be called Lyapunov function following Prigogine.  In this way, we rename the velocity potential $\phi$ as entropy $S=\phi$,
the existence of which is guaranteed by the Helmholtz (or the Hodge) theorem.\footnote{
 In terms of hydrodynamics, the number density $\rho(t, \bm{x})/m~(m$: mass of fluid particle) is the distribution function $f(t, \bm{x})$ in the $N$-dimensional phase space.  Here, we consider $m=1$ and identify $f= \rho \equiv n$.  Then, the number density $n(t, \bm{x})$ of fluid particle changes in the compressible fluid, following the equation of continuity, 
\begin{eqnarray}
D_t \; n(t, \bm{x})=\left(\partial_t + \bm{v} \cdot \bm{\nabla} \right) n(t, \bm{x})= - (\bm{\nabla} \bm{v}) \; n(t, \bm{x}), 
\end{eqnarray}
where $D_t$ is the Lagrangian derivative. In case of the compressible fluid ($\bm{\nabla} \bm{v} \ne 0$), the phase space is not uniformly distributed, that is $f=n$ is not constant, but the concentration or diffusion occurs within the phase space while streaming. This can be considered as the Lyapunov effect. Further details are continued in Appendix D. }

Then, the second term is written by

\begin{equation}
\bm{v}^{(2)}(\bm{x})=\bm{\nabla}S.
\end{equation}

In terms of the Hamiltonians $H_{1},\dots,H_{N-1}$ and the entropy $S$, the equation of motion of our system now becomes
\begin{equation}
\frac{d\bm{x}}{dt}=\{\bm{x},H_{1},H_{2},\dots,H_{N-1}\}_{NB}+\bm{\nabla}S, \label{Nambu thermo 1}
\end{equation}
and for a general operator $O(x)$, we have
\begin{equation}
\dot{O}=\dot{\bm{x}}\cdot\bm{\nabla}O=\{O,H_{1},H_{2},\dots,H_{N-1}\}_{NB}+\bm{\nabla}O\cdot\bm{\nabla}S. \label{Nambu thermo 2}
\end{equation}

The first term in the right-hand side is the Nambu dynamics, a generalization
of the Hamilton dynamics to $N$-dimensional phase space or the fluid
dynamics in $N$-dimensional thermodynamic space without dissipation.
The second term is relevant to the dissipation in the non-equilibrium
thermodynamics involving the entropy.  

Accordingly, we call this theory as ``Nambu
non-equilibrium thermodynamics''.\footnote{
We can consider closed and open systems because the inflow
and outflow of entropy and materials can be introduced as advection terms in the Nambu dynamics. We also consider isolated systems without such inflows and outflows.
}

What we have shown here is far from the ``existence theorem'' for Nambu non-equilibriun thermodyamics.  

We were aiming to show the following: For any autonomous differential equations with $N$ variables, we could find $N-1$ Hamiltonians and one entropy $S$, so that the autonomous differential equations could be written by Eqs.(\ref{Nambu thermo 1}) and (\ref{Nambu thermo 2}).

As was stated above, although the existence theorem exists \underline{locally}, it can not be proved \underline{globally}.  Therefore, we have to assume the existence of $H_1-H_{N-1}$ from the beginning.

In the process of proving the existence theorem,  however, we can recognize an interesting problem of how the breakdown of global existence of Nambu Hamiltonians is related to the chaotic or singular behavior of the system. This is beyond the scope of our paper, and is a future issue.  Even under the assumption of existence of Hamiltonians, we can discuss a very interesting behavior of cycles giving spike-like signal propagations. In this paper we will study some of these issues. 

\color{black}

\subsection{LCR circuit}

A typical dynamical system with
dissipation\ is the LCR circuit.\footnote{It is well known that LCR circuits can be described as dynamical systems.
See for example \cite{key-43}.}. 

In an LCR circuit, the voltage can be written as the sum of the
contributions from resistance $R$, inductance $L$ and capacitance
$C$,

\begin{equation}
V=V_{R}+V_{L}+V_{C}
\end{equation}
where

\begin{equation}
V_{R}=IR,\ V_{L}=L\dot{I},\ V_{C}=\frac{Q}{C},
\end{equation}
and $I$ is electric current and $Q$ is electric charge with $I=\dot{Q}$, and 
$\Phi$ is the magnetic flux inside the coil with $\Phi=LI$.

We can express the LCR circuit, in terms of Nambu non-equilibriun thermodynamics, 
\begin{equation}
\dot{Q}=\frac{\partial H}{\partial\Phi}-\frac{\partial S}{\partial Q}, ~~
\dot{\Phi}=-\frac{\partial H}{\partial Q}-\frac{\partial S}{\partial\Phi}.
\end{equation}
with one Hamiltonian and one entropy,
\begin{equation}
H=\frac{1}{2L}\Phi^{2}+\frac{1}{2C}Q^{2}\label{eq:H}, ~~S=\frac{R\Phi^{2}}{2L}.
\end{equation}
Or, denoting $Q$ and $\Phi$ as $x^{1}=Q,\ x^{2}=\Phi$, then we have 
\begin{equation}
\dot{x}^{i}=-\{H,x^{i}\}_{PB}+\{x^{i},x^{j}\}_{PB}\{S,x_{j}\}_{PB}.
\end{equation}

Here we examine the time evolution of entropy and Hamiltonian in this simple example, which gives
\begin{equation}
\dot{S}=-\{H,S\}_{PB}+\{S,x_{i}\}\{S,x^{i}\}_{PB},
\end{equation}

\begin{equation}
\dot{H}=\{H,x_{j}\}_{PB}\{S,x^{j}\}_{PB}.
\end{equation}

Generalizing the foregoing to Nambu non-equilibrium thermodynamics,
the time evolution of entropy is 

\begin{equation}
\dot{S}=-\{H_{1},\dots,H_{n-1},S\}_{NB}+\{S,x_{i_{1}},\dots,x_{i_{n-1}}\}_{NB}\{x^{i_{1}},\dots,x^{i_{n-1}},S\}_{NB},
\end{equation}

\begin{equation}
\dot{H}_{i}=\{H_{i},x_{i_{1}},\dots,x_{i_{n-1}}\}_{NB}\{x^{i_{1}},\dots.,x^{i_{n-1}},S\}_{NB}.
\end{equation}

\color{black}
What should be noted here is that ``the time evolution can be divided into two contributions'', the Hamiltonian part contribution and the entropy part contribution.  The division of the two contributions provides a good information to understand what is happening in the time evolution. We will denote this division by 
\begin{align}
    \dot{O} &= \partial^{(H)}_t O + \partial^{(S)}_t O, \\
    \partial_t^{(H)} O &\equiv -\{H_1,\dots,H_{n-1},O\}_{NB}, \\
    \partial_t^{(S)} O &\equiv \{S,x_1,\dots,x_{n-1}\}_{NB}\{x^1,\dots,x^{n-1},O\}_{NB}.
\end{align}

The entropy $S$ introduced in the last and this section corresponds to thermodynamic entropy, which is discussed in Appendix A.

\subsection{Principle of minimum dissipation of entropy}

Let us introduce fluctuating entropy as in Onsager's discussion of non-equilibrium thermodynamics\footnote{In the following description, subscripts of NB are omitted.}. 

\color{black}
Let us simply add a thermal fluctuation term to the time evolution,
\begin{equation}\label{eq:fluctuation}
\dot{x}^{i}= -\{H_1,H_2,x^i\}+\frac{\partial S}{\partial x^i}+\xi^{i}(t),
\end{equation}
where $\xi^{i}(t)$ is a Gaussian white noise with zero mean and variance,
\color{black}
\color{black}

From the ordinary discussion of the case of adding Gaussian fluctuations,
\footnote{
  \color{black}
  The probability density of the path $x(t)$ of a stochastic process is 
  \begin{equation}
    P(x)\propto  \int D\xi \exp \left(- \int \frac{1}{2} \xi (t)^2 \right) 
  \end{equation}
   if Gaussian fluctuations $\xi^i$ are used. The fluctuation is related to the thermodynamic variable $x(t)$ through the constitutional equation (\ref{eq:fluctuation}). 
   Therefore, the Gaussian distribution is converted to the distribution for a stochastically created path $x(t)$, leading to the Onsager-Machlup-Hashitsume action $S_{OMH}$\cite{key-100}\cite{key-101}\cite{key-102}\cite{key-103}.
   Note that this action drops the correction derived from the Jacobian due to variable transformation.
   
 Since this action gives the probability distribution of the stochastic process, it is relevant to the entropy of Shanon or Boltzmann. }  we obtain
\color{black}
\begin{align}
    \dot{S}_{OMH} &\equiv \Phi -\partial^{(S)}_t S +\Psi \\
    &=\frac{1}{2} \left(\dot{x}^i + \{H_1,H_2,x^i\}-\frac{\partial S}{\partial x^i}\right) \left(\dot{x}^i + \{H_1,H_2,x^i\}-\frac{\partial S}{\partial x^i}\right)
\end{align}
where 
\begin{align}
    \Phi&\equiv\frac{1}{2}\left(\dot{x}_{i}-\{H_{1},H_{2},x_{i}\}\right)^{2}, \\
    \Psi&=\frac{1}{2}\frac{\partial S}{\partial x_{i}}\frac{\partial S}{\partial x_{i}}.
\end{align}

Assuming the principle of minimum dissipation of entropy $\dot{S}_{\text{OMH}}=0$ as in Onsager's argument, we obtain 
\begin{equation}
    \partial_{t}^{(S)}S=\Phi+\Psi\geq0
\end{equation} 

This means $\partial^{(S)}_t S \ge 0$, that is, the contribution of time evolution with respect to entropy, is always positive.

It should be noted that the total time evolution of the entropy includes the Hamiltonian contribution 
\begin{equation}
    \dot{S} = \partial_t^{(H)} S + \partial_t^{(S)} S,
\end{equation}
so that $\dot{S}$ can be negative. This is a necessary condition for the path to cycle.  It is because, when a cycle appears, the entropy also comes back to the original value after a single tern of the cycle, for which to occur, both the increasing period and the decreasing period for the entropy are necessary.  The latter decrease can be afforded through the evolution by the Hamiltonians in the Nambu dynamics. 

\subsection{Cycle}
In this subsection, we discuss the triangular reaction far from equilibrium, in terms of Nambu non-equilibrium thermodynamics.

Such a system does not satisfy the detailed balance, where the reaction equation has only the right-handed cycle without the left-handed one, as follows.
\begin{equation}
\xymatrix{X_{1}\ar[r]^{k_{12}} & X_{2}\ar[ld]^{k_{23}}\\
X_{3}\ar[u]^{k_{31}}
}
\end{equation}
The time evolution of concentrations is as follows.
\begin{align}
\frac{dx_{1}}{dt}=k_{12}x_{1}-k_{31}x_{3} \\
\frac{dx_{2}}{dt}=k_{23}x_{2}-k_{12}x_{1} \\
\frac{dx_{3}}{dt}=k_{31}x_{3}-k_{23}x_{2}
\end{align}

In such a situation, the time evolution is not proportional to the thermodynamic force, but is non-linearly related to the force \footnote{Here, $x_i=e^{\beta \Delta \mu_i}$}.  \color{black}

Furthermore, the constants of reaction rate $k_{ij}$ is not symmetric but antisymmetric, so that Onsager's theory cannot be applied to the time evolution. On the other hand, Nambu non-equilibrium thermodynamics can handle it.

More specifically, we can discuss an example by chooseing the following Hamiltonians $H_1,H_2$ and ``entropy'' $S$,

\begin{equation}
    H_{1}=\frac{1}{4}(x_{1}^{2}+x_{2}^{2}+x_{3}^{2}),
\end{equation}
\begin{equation}
    H_{2}=k_{23}x_{1}+k_{31}x_{2}+k_{12}x_{3},
\end{equation}
\begin{equation}
    S=-\frac{1}{2}(k_{12}x_{1}^{2}+k_{23}x_{2}^{2}+k_{31}x_{3}^{2})+\frac{1}{2}(k_{12}x_{1}x_{2}+k_{23}x_{2}x_{3}+k_{31}x_{3}x_{1}).
\end{equation}

The system is notable for the following conservation law to hold,

\begin{equation}
    \frac{d}{dt} \left(x_1 + x_2 + x_3 \right) = 0
\end{equation}
which indicates that the thermodynamic variables continue to move, while satisfying always the conservation law, so that the system never settle into a single state. In other words, it does not move toward a state of thermodynamic equilibrium.  This is an example of far from the thermal equilibrium.

The more general triangular reaction, including the non-linearity and the cycle, can also be described by Nambu non-equilibrium thermodynamics, as follows
\begin{equation}
    H_{1}=\frac{1}{4}(x_{1}^{2}+x_{2}^{2}+x_{3}^{2}),
\end{equation}
\begin{align}
    H_{2}=k_{23}\left(1-\frac{k_{32}}{k_{23}}e^{\beta(A_{23}-A_{(0)23})}\right)x_{1} \nonumber \\
+k_{31}\left(1-\frac{k_{13}}{k_{31}}e^{\beta(A_{31}-A_{(0)31})}\right)x_{2} \nonumber \\
+k_{12}\left(1-\frac{k_{21}}{k_{12}}e^{\beta(A_{12}-A_{(0)12})}\right)x_{3},
\end{align}
\begin{align}
    S=-\frac{1}{2}\left[k_{12}\left(1-\frac{k_{21}}{k_{12}}e^{\beta(A_{12}-A_{(0)12})}\right)\left(x_{1}^{2}-x_{1}x_{2}\right)\right. \nonumber \\
+k_{23}\left(1-\frac{k_{32}}{k_{23}}e^{\beta(A_{23}-A_{(0)23})}\right)\left(x_{2}^{2}-x_{2}x_{3}\right) \nonumber \\
+\left.k_{31}\left(1-\frac{k_{13}}{k_{31}}e^{\beta(A_{31}-A_{(0)31})}\right)\left(x_{3}^{2}-x_{3}x_{1}\right)\right].
\end{align}

In this way, we can show the examples far from equilibrium, having cycles, which can be described by Nambu non-equilibrium thermodynamics.

\subsection{Summary of Nambu non-equilibrium thermodynamics}
Let us summarize the important points we have discussed so far. First, while Onsager's method can be applied to triangular reactions close to equilibrium, as in Section 2.3, it cannot be applied to non-equilibrium states that are extremely far from equilibrium, such as those with cycles, as described in Section 2.7.

Second, Onsager's method has the limitation that the principle of dynamic reversibility is a requirement, which fails to describe the systems with cycles, as was discussed in section 2.7. 

This limitation can be overcome by introducing a Nambu bracket. It is because, in the absence of dissipative terms, the Nambu dynamics is given by the Nambu bracket, having multiple conserved quantities, Hamiltonians. This fact guarantees the description of cycles. After introducing the dissipative terms, we arrive at the Nambu non-equilibrium thermodynamics.

\color{black}We could not established completely, the ``existence theorem'' which states that any autonomous system for $N$ variables, the $N-1$ Hamiltonians $\{H_1, \cdots, H_{N-1}\}$ and one entropy $S$, such that the original the set of autonomous differential equations can be expressed in terms of the Nambu non-equilibrium thermodynamics.
Although the ``existence theorem'' can be proved locally, there exists a lot of counter-examples for the global existence theorem.  Therefore, we have introduced the Nambu non-equilibrium thermodynamics under the assumption of the existence of $N-1$ Hamiltonians. \color{black} 

The entropy terms may break some of the conservation laws, existed originally in the Nambu dynamics. However, if one of the conserved quantities remains, the thermodynamics can describe the cycle. Even if all the conservation laws are broken, there is a possibility for the cyclic motion to remain, for which we call the system has ``a pseudo-conservation law ''\footnote{Relationships that can be regarded as approximately conservative on a given time scale.}. This may happen from the competition between two ``motive forces''; one is to make cycles by Nambu dynamics with multiple Hamiltonians, and the other is the dissipation by entropy. For the cyclic motion (cycle) to occur, we have the increasing period of entropy and the decreasing period of it. This oscillatory behavior may be realized under a certain competition of the above mentioned two motive forces. The problem we study in this paper is, not limited to the triangular reactions, but are applicable to the vast non-equilibrium phenomena far from the equilibrium, in particular for the systems with limit cycles, in which Nambu non-equilibrium thermodynamics plays as important role in understanding the phenomena. Some examples to support our claim are given, in the LCR circuit, the chemical BZ reactions, and the signal propagation in the nerve.

\color{black}
The theory developed by M. Grmela and H.C. \"Ottinge \cite{key-50,key-200} overlaps  with our paper, only when we restrict our Nambu non-equilibrium thermodynamics to a special case with a single Hamiltonian and the linear response near the equilibrium. The main interest of ours is, however, the systems far from the equilibrium, which are not covered by \cite{key-201, key-202}.  On the other hand, Nambu non-equilibrium thermodynamics developed in this paper can analyze them, since  it consists of multi-Hamiltonian Nambu dynamics as well as of the non-equilibrium thermodynamics introduced naturally through the incompressibility.
\color{black}

\color{black}
\section{Derivation of the general Form for Nambu non-equilibrium thermodynamics}

In Nambu non-equilibrium thermodynamics, entropy plays a crucial role
in understanding the behavior of systems far from equilibrium. The
entropy change $dS$ can generally be divided into two components:
the part related to the inflow and outflow of entropy, and the part corresponding to the entropy generation
within the system itself. From the previous discussions on the Nambu non-equilibrium thermodynamics, we can identify the first component with $d^{(H)}S$, and the second one to  $d^{(S)}S$:

\begin{equation}
dS=d^{(H)}S+d^{(S)}S.
\end{equation}

This separation allows for a clearer analysis of the system's evolution
and the factors contributing to its entropy changes.

The first term, $d^{(H)}S$, represents the exchange of entropy between
the system and its surrounding environment, which occurs through
various processes such as heat flow, mass transfer, and radiation.
This term is important when considering the interaction of the system
with its environment and how it affects the overall entropy balance.

The second term, $d^{(S)}S$, accounts for the internal entropy generation
within the system. This term arises from irreversible processes, such
as chemical reactions, dissipation, and relaxation phenomena, which
lead to an increase in the system's entropy. By analyzing $d^{(S)}S$,
we can gain insights into the driving forces behind the system's evolution
and the factors that contribute to its irreversibility. From the second
law of thermodynamics, $d^{(S)}S$ always increases,
\begin{equation}
d^{(S)}S\geq0.
\end{equation}
Since $dS$ is a total derivative, taking the integral around the
circle yields 

\begin{equation}
0=\oint dS=\oint\left(d^{(H)}S+d^{(S)}S\right),
\end{equation}
so 
\begin{equation}
\color{black}-\oint d^{(S)}S=\oint d^{(H)}S\equiv\oint\frac{dQ}{T}\color{black}
\end{equation}
where $dQ$ is the heat energy flowing into the system, and
$T$ is the temperature of the system

\footnote{\color{black} The cycle describes a cyclic heat engine, for which we separate the infinitesimal heat flows $dQ_i~(i=1, 2, \cdots)$ into positive (coming-in) ones $P$ and negative (going-out) ones $N$, and introduce the \emph{averaged high temperature} $1/T_1=\langle 1/T \rangle_P$ and the \emph{averaged low temperature} $1/T_2=\langle 1/T \rangle_N$.  Here, the average means $\langle 1/T \rangle_{P(N)}= \sum_{i \in P(N)} (1/T) |dQ_i|  / Q_1(Q_2)$, where $Q_1=\sum_{i \in P}  dQ_i$ and $Q_2= -\sum_{i \in N}  dQ_i$ are chosen positively.

Then, we have for any cyclic engine
$ \oint \frac{dQ}{T} = \frac{Q_1}{T_1} - \frac{Q_2}{T_2}$, similar to the Carnot cycle. The energy conservation $Q_1 - Q_2=W$ with $W$ the work done, and the efficiency $\eta=\frac{W}{Q_1}$, we obtain
$
\oint \frac{dQ}{T}  =  \frac{Q_1}{T_2}\left(\eta- \frac{T_1-T_2}{T_1} \right).$ 
Now from (3.4), we find that $$
\left[ \oint d^{(S)} S \ge 0\right]  \Leftrightarrow \left[ \eta \le \frac{T_1-T_2}{T_1},  ~\text{or}~ T_1 \ge T_2 ~~\text{for}~W=0 \right].$$
This shows the implication of entropy production $\oint d^{(S)} S \ge 0$, if we consider $1/T_1$ and $1/T_2$ are the \emph{averages} over the contribution of heats, coming-in and going-out, respectively.  As is learnt from the case of $W=0$, the entropy production is consistent with the nature of heat that it flows from the higher temperature parts (in average) to the lower temperature pars (in average). \color{black}}
\footnote{
  \color{black}
  Whether there is a \(d^{(H)}S\) contribution or not depends on whether the system is isolated or not. 
  
  In an isolated system, there is no entropy exchange with the environment (\(d^{(H)}S = 0\)), and the entropy change is solely due to internal irreversibility (\(d^{(S)}S\)). As a result, the cycle condition \(\oint dS = 0\) cannot generally restore the entropy to its original state because the internal entropy generation (\(d^{(S)}S \geq 0\)) always increases the total entropy in irreversible processes.

  However, \(\oint dS = 0\) can still hold in an isolated system if and only if the internal entropy generation vanishes (\(d^{(S)}S = 0\)), which corresponds to a completely reversible process. This highlights that \(\oint dS = 0\) is a special case in isolated systems and cannot be generally assumed.
  \color{black}
}
.

On the other hand, because $d^{(H)}S$ can be expanded like 

\begin{equation}
d^{(H)}S=\frac{\partial S}{\partial U}d^{(H)}U+\frac{\partial S}{\partial V}d^{(H)}V+\frac{\partial S}{\partial M}d^{(H)}M+\dots
\end{equation}

\begin{equation}
=\frac{1}{T}d^{(H)}U-\frac{p}{T}d^{(H)}V+\frac{\mu}{T}d^{(H)}M+\dots
\end{equation}

\begin{equation}
\equiv\frac{\partial S}{\partial x^{i}}d^{(H)}x^{i},
\end{equation}
where $U$ is the internal energy, $V$ is the volume, $M$ is the
number of particles, $p$ is the pressure, and $\mu$ is the chemical
potential, and $x^{i}$, $i=1,\dots,N$, represents any of the variables
$U,V,M,\dots$.

Similarly,

\begin{equation}
d^{(S)}S=\frac{\partial S}{\partial x^{i}}d^{(S)}x^{i}.
\end{equation}

From above,
\begin{equation}
\frac{dS}{dt}=\frac{\partial S}{\partial x^{i}}\frac{d^{(H)}x^{i}}{dt}+\frac{\partial S}{\partial x^{i}}\frac{d^{(S)}x^{i}}{dt}.
\end{equation}

We may consider the first term $d^{(H)}S/dt$ as the change of entropy by a reversible process, while the second term $d^{(S)}S/dt$ be that by an irrreversible process, since the first term implies the inflow and outflow of entropy, while the second gives the production of entropy. 

Accordingly, we need to have 
\begin{equation}
\frac{d^{(H)}S}{dt}=\frac{\partial S}{\partial x^{i}}\frac{d^{(H)}x^{i}}{dt}=0.
\end{equation}
\color{black}when the system is reversible and isolated. In the reversible process $d^{(S)}S/dt = 0$, but in the irreversible process $d^{(S)}S/dt > 0$.\color{black}

To satisfy this condition, a sufficient condition is to introduce
the antisymmetric tensor $B_{ij}(x)$, and to write the current,  $d^{(H)}x^{i}/dt$ in the following form 
\begin{equation}
\frac{d^{(H)}x^{i}}{dt}=B^{ij}\frac{\partial S}{\partial x^{j}}.
\end{equation}

Written in this way, the time evolution of the reversible part of
entropy is shown to vanish, since

\begin{equation}
  \frac{d^{(H)}S}{dt}=B^{ij}(x)\frac{\partial S}{\partial x^{i}}\frac{\partial S}{\partial x^{j}}\equiv\{S,S\}_{B}=0. \label{reversible part of current}
\end{equation}

This guarantees vanishing of the time evolution of the reversible part of entropy, which is consistent with the second law of thermodynamics.

In other words, the entropy of a closed system undergoing reversible
changes remains constant over time, while irreversible changes lead
to an increase in entropy.

The bracket $\{S,O\}_{B}$ can be reproduced by using the Nambu dynamics.

It is because, if we choose $B_{ij}(x)$ as,
\begin{equation}
B^{ij}(x) \equiv \color{black}-\color{black}\epsilon^{i_{1}\dots,i_{N-2}ij}\frac{\partial H_{1}}{\partial x^{i_{1}}}\dots\frac{\partial H_{N-2}}{\partial x^{i_{N-2}}},
\end{equation}
we obtain the bracket $\{S,O\}_{B}$ in the Nambu dynamics as
\begin{equation}
\{H_{1},\dots,H_{N-2},S,O\}_{NB}\equiv\{S,O\}_{B} \label{eq:qualconditionH}.
\end{equation}

\subsubsection*{$\circ$ Non-equilibrium process close to equilibrium; linear response}
Next, we consider the irreversible process.
The time evolution of the thermodynamic variables is

\begin{equation}
\frac{d^{(S)}x^{i}}{dt}=L^{ij}\frac{\partial S}{\partial x^{j}}\equiv L^{ij}A_{j},
\end{equation}
where $L^{ij}$ is transport coefficient and is a second-order symmetric
tensor and $A_{i}$ is the affinity force.\footnote{Usually the kinetic constant $L^{ij}$ is a constant, but as $B^{ij}(x)$ depends on the coordinate, $L^{ij}$ can be a function of the coordinate $x^i$.  Then, $L^{ij}(x)$ and $R_{ij}(x)$ are regarded as the metric tensor $g^{ij}(x)$ and $g_{ij}(x)$ of the space spanned by the thermodynamic variables $(x^1, \cdots, x^{N})$.}

Here, we have to compare this expression to that in the reversible process Eq.(\ref{reversible part of current}), which is
\begin{equation}
\frac{d^{(H)}x^{i}}{dt}=B^{ij}\frac{\partial S}{\partial x^{j}} = B^{ij}A_{j}.
\end{equation}

The symmetric $L^{ij}=L^{ji}$ and the anti-symmetric $B^{ij}=-B^{ji}$ separate the irreversible process with entropy production from the reversible process by the ordinary dynamics.
As in the reversible case, in the irreversible
case, we will use another bracket notation as well,

\begin{equation}
\frac{d^{(S)}O}{dt}=L^{ij}\frac{\partial S}{\partial x^{i}}\frac{\partial O}{\partial x^{j}}\equiv(S,O)_{L_{2}},
\end{equation}

\begin{equation}
(x^{i},x^{j})_{L_{2}}=L^{ij}.
\end{equation}

In summary, the thermodynamics of non-equilibrium systems close to
equilibrium can be described in terms of Nambu Non-Equilibrium Thermodynamics as follows:

\begin{equation}\label{eq:close to equilibrium}
\frac{dO}{dt}=\color{black}-\color{black}\{H_{1},\dots,H_{N-2},S,O\}_{NB}+(S,O)_{L_{2}},
\end{equation}

\begin{equation}
\frac{dS}{dt}=(S,S)_{L_{2}}.
\end{equation}

\subsubsection*{$\circ$ Non-equilibrium process far from equilibrium; non-linear response}

The most important point, as shown below, is that the Nambu non-equilibrium
thermodynamics description is applicable even to systems far from
equilibrium.

In such systems, the equations of motion or the constitutive equations, are not linear with respect
to the affinity force, and in general higher-order transport coefficients
appear.

Therefore, this case is described by the following general form,

\begin{equation}
\frac{dO}{dt}=\color{black}-\color{black}\{H_{1},\dots,H_{N-2},S,O\}_{NB}+\sum_{i=0}^{\infty}(\stackrel{i+1}{\overbrace{S,\dots,S}},O)_{L_{i+2}}.
\end{equation}
The notation introduced in the above represents the non-linear response, where $(i+1)$ product of affinity forces $A_a(x)$ with a kinetic constant $L^{a_1 \cdots a_{j+1}}$ gives the ``thermodynamic current''. 
\begin{eqnarray}
&& (\stackrel{i+1}{\overbrace{S,\dots,S}},O)_{L_{i+2}}  \equiv \frac{d^{(S)}x^{a_{i+2}}}{dt} \frac{\partial O}{\partial x^{a_{i+2}}}, ~\text{with} ~A_a(x)=\frac{\partial S}{\partial x^a}, ~\text{and} \\
&&\frac{d^{(S)}x^{a_{i+2}}}{dt}= \sum_{a_1,a_2, \cdots, a_{i+1} }L^{a_1a_2 \cdots a_{i+1}a_{i+2} } A_{a_1}(x) A_{a_2}(x) \cdots A_{a_{i+1}}(x),~~~~~
\end{eqnarray}
where $L^{a_1a_2 \cdots a_{i+1} a_{i+2} }$ are $(i+1)$-order complete symmetric tensors.

Here we make an important assumption: the system we are considering can include ``open systems'', where the system does not necessarily go
to equilibrium. Therefore, we will more generally write the reversible
part, by using a general $H_{N-1}$ instead of $S$, as follows:
\begin{equation}
\frac{dO}{dt}=\color{black}-\color{black}\{H_{1},\dots,H_{N-2},H_{N-1},O\}_{NB}+\dots.
\end{equation}

The important point for this choice is that $d^{(H)} x^i/dt$ takes the usual form of the Nambu dynamics,
\begin{eqnarray}
\frac{d^{(H)}x^i}{dt}= B^{ij}(x) \frac{\partial H_{N-1}}{\partial x^j} = \color{black}-\color{black}\{H_1, \cdots, H_{N-1}, x^i \}_{NB},
\end{eqnarray}
so that $d^{(H)}S/dt \ne 0$, which is, however,  favorable in the open system, where the excess of the inflow entropy is absorbed by the environment 
\footnote{\color{black}
This is the crucial difference between systems close to equilibrium and those far from equilibrium. In systems close to equilibrium, as discussed above in (\ref{eq:close to equilibrium}), one of the Hamiltonians is S and $d^{(H)}S/dt=0$.
\color{black}
}.

\subsubsection*{$\circ$ Coarse-graining and appearance of cycle}
Next, we discuss a possibility that in such a non-linear system, the system does not move toward the equilibrium, but rather, approaches a ``cyclic system'' (denoted by $\Omega_{\mathrm{cycle}}$) in the Nambu non-equilibrium thermodynamics.  

We will explain this in the following.

In the Nambu non-equilibrium thermodynamics, the general non-linear response can be described by
\begin{eqnarray}
\dot{x}^{j}= \color{black}-\color{black}\{ H_1, H_2, \cdots, H_{N-1}, x^{j} \}_{NB} +  
\sum_{i=0}^{\infty} (\stackrel{i+1}{\overbrace{S,\dots,S}}, x^{j})_{L_{i+2}}.
\end{eqnarray}
The second term in the right-hand side of the above equation has a complicated expression with the multiple product of $\partial S/ \partial x^{a}$, but is nothing but a vector field.  As was discussed in Section 2, we know that any vector field can be expressed in terms of the Nambu non-equilibrium thermodynamics. \color{black}This naive understanding is not rigorously correct, but we are allowed to consider it valid in the normal cases. \color{black}Therefore, we can find $\{H'_1, \cdot, H'_{N-1} \}$ and a new entropy $S^C$, and obtain the following expression:
\begin{eqnarray}
\sum_{i=0}^{\infty} (\stackrel{i+1}{\overbrace{S,\dots,S}}, x^{j})_{L_{i+2}}=  \color{black}-\color{black}\{ H'_1, H'_2, \cdots, H'_{N-1}, x^{j} \}_{NB} + \partial_j S^C.
\end{eqnarray}

Thus, we have 
\begin{eqnarray}
\dot{x}^{j}= \color{black}-\color{black}\{ H_1, H_2, \cdots, H_{N-1}, x^{j} \}_{NB} \color{black}-\color{black}  \{ H'_1, H'_2, \cdots, H'_{N-1}, x^{j} \}_{NB} + \partial_j S^C.
\end{eqnarray}
Looking at the first and second terms in the above right-hand side, each satisfies the incompressibility condition, so that the sum of them also satisfies the condition.  Thus, we can find the third set of Hamiltonians $\{ H^C_{1}, H^C_{2}, \cdots, H^C_{N-1} \}$, which yields the following general expression, even in the case far from equilibrium,
\begin{eqnarray}
\dot{x}^{j}= \color{black}-\color{black}\{ H^C_1, H^C_2, \cdots, H^C_{N-1}, x^{j} \}_{NB} + \partial_j S^C.
\label{xtime}
\end{eqnarray}
For a general observable $O$, we have
\begin{eqnarray}
\dot{O}= \color{black}-\color{black}\{ H^C_1, H^C_2, \cdots, H^C_{N-1}, O \}_{NB} + (\nabla S^C \cdot \nabla O).
\label{otime}
\end{eqnarray}

The non-linear response again gives the system a simpler form, such as the non-equilibrium thermodynamics near Onsager's equilibrium system. It is described by a third entropy $S^C$, different from Onsager's entropy S, which gives $dS=dQ/T$ and the Shannon entropy $S_{OMH}$. This drives the system into a circular system $\Omega^C$, i.e., satisfying the following conditions
\begin{eqnarray}
\frac{d S^C}{dt} \ge 0, ~\text{and}~\frac{d^2 S^C}{dt^2} < 0.
\end{eqnarray} 

Under this condition, after a certain period, the system is relaxed to the cyclic system $\Omega^C$, which is again controlled by a pure Nambu dynamics,
\begin{eqnarray}
\dot{O}= \color{black}-\color{black}\{ H^C_1, H^C_2, \cdots, H^C_{N-1}, O \}_{NB}.
\end{eqnarray}

Then, the conservation laws of $H^C_i=$ constant are recovered and the cyclic motion arises. 

This is nothing but a guess, not being able to prove in this paper. Probably because Boltzmann's approach averages the dynamical system over small periods, the dynamics of thermodynamics may be coarse-grained on larger time scales. Thus, the coarse-graining of Nambu non-equilibrium thermodynamics may be worth considering.
Our guess may be derived from such coarse-graining.  
                                                              
\section{Approaches for constructing the Hamiltonians and entropy}

We start from the general form of the Nambu non-equilibrium thermodynamics (\ref{xtime}) and (\ref{otime}), given in the last section, which is applicable to the system far from the equilibrium with non-linear response and cycles. 
They are 
\begin{eqnarray}
&&\dot{x}^{i}=\color{black}-\color{black}\{ H^C_1, H^C_2, \cdots, H^C_{N-1}, x^{j} \}_{NB} + \partial_i S^C,
\end{eqnarray}
and for $O$
\begin{eqnarray}
&&\dot{O}=\color{black}-\color{black} \{ H^C_1, H^C_2, \cdots, H^C_{N-1}, O \}_{NB} + (\nabla S^C \cdot \nabla O).
\end{eqnarray}

In comparison with Subsection 1.4, these equations reproduce the set of $N$ autonomous differential equations,
\begin{eqnarray}
\dot{x}^i= v^C(x)^i~~(i=1, 2, \cdots, N).
\end{eqnarray}

The entropy is easier to find, since it satisfies $\nabla \cdot\bm{v}(x)= \nabla^2 S^C(x)$.

Even for a general $N$, we may find the solution for the Hamiltonians and entropy \color{black}in the normal cases, \color{black}but here, we restrict to $N=3$, since all the examples we study belong to this case.  Then, we have two Hamiltonians, $H^C_1$ and $H^C_2$. We may denote $\bm{x}=(X, Y, Z)$.

The velocity field is the sum of the incompressible part, by Nambu dynamics with conserved stream functions or Hamiltonians, and the compressible part, due to entropy.  For $N=3$, this is
\begin{eqnarray}
\bm{v}(x)= \nabla \times \bm{ \psi}+ \nabla S^C,
\end{eqnarray}
from which we obtain
\begin{eqnarray}
\bm{\psi}=  \frac{1}{2}  (H^C_1 \overleftrightarrow{\nabla} H^C_2).
\end{eqnarray}

An easy way to find the solution is to impose
\begin{equation}
H^C_{2}=Z,
\end{equation}
which works for a number of examples which we will study. It is, however, an open problem whether we can impose this condition in any case, by using the arbitrariness existing possibly in the choice of Hamiltonians and entropy.

Under this assumption, $H^C_1$ and $S^C$ can be obtained, directly from the autonomous differential equations, 
\begin{eqnarray}
\begin{cases}
\; \dot{x}^1=(v^C)^1(x)=  2 \partial_2 H_1^C + \partial_1 S^C, \\
\; \dot{x}^2=(v^C)^2(x)=-2 \partial_1 H_1^C + \partial_2 S^C, \\
\; \dot{x}^3=(v^C)^3(x)=\partial_3 S^C,
\end{cases}
\end{eqnarray}
where we may use $(x^1, x^2, x^3)=(X, Y, Z)$ in the following.

Combining these equations yields the following relational equations:

\begin{equation}
\begin{aligned}
&2\partial_{2}^{2}H^C_{1}+2\partial_{1}^{2}H^C_{1}&=\partial_{2}\dot{x}^{1}-\partial_{1}\dot{x}^{2}, \\
&2\partial_{3}\partial_{1}H^C_{1}&=\partial_{3}\dot{x}^{1}-\partial_{1}\dot{x}^{3}, \\
&-2\partial_{3}\partial_{2}H^C_{1}&=\partial_{3}\dot{x}^{2}-\partial_{2}\dot{x}^{3}.
\label{HSCondition}
\end{aligned}
\end{equation}

These relational equations are also derived from another discussions.

The constitutive equation as a starting point for the discussion is as follows\footnote{In the following, $C$ superscripts are omitted as appropriate.}:
\begin{equation}
\dot{x}^i = v^i = \color{black}-\color{black} \{H_1,H_2,\dots, H_{N-2},H_{N-1},x^i\}_{NB} + \partial_i 
  S.
\end{equation}

In Nambu non-equilibrium thermodynamics, the constitutive equation
can be rewritten in the following differential formalism

\begin{equation}
dS_{NB}=\delta_{ij}\left(v^{i}+\{H_{1},\dots,H_{N-2},H_{N-1},x^{i}\}_{NB}\right)dx^j.
\end{equation}

Therefore, by taking the exterior derivative of both sides, the following
relationship can be obtained,

\begin{equation}
\begin{aligned}\delta_{ij}\left(\frac{\partial v^{j}}{\partial x^{k}}+\frac{\partial}{\partial x^{k}}\{H_{1},\dots,H_{N-2},H_{N-1},x^{j}\}_{NB}\right)dx^{k}\wedge dx^{i} & =0\end{aligned}
\label{eq:relation}
\end{equation}

Therefore, if the Hamiltonian is introduced to satisfy this condition,
the remaining terms in the constitutive equation can always be written
in terms of the gradient of entropy from the condition of exact differential.

Now consider a chemical reaction system when there are three dynamical
variables $X,Y,Z$. Given a chemical species $Z$, which is the catalyst
in the chemical reaction here, it is natural that its concentration
will cycle. Therefore, it is one of two Hamiltonians,

\begin{equation}
H_{2}=Z.
\end{equation}

In this case, from the relation (\ref{eq:relation}), we obtain

\begin{equation}
\left(\delta_{kj}\frac{\partial^{2}H_{1}}{\partial x^{i}\partial x^{j}}+\delta_{ij}\frac{\partial v^{j}}{\partial x^{k}}\right)dx^{k}\wedge dx^{i}=0.
\end{equation}

This corresponds to (\ref{HSCondition}).

We will confirm later with the BZ reaction and the Hindmarsh-Rose
model that $H_{1}$ can indeed be found to satisfy this relationship.

\color{black}

\section{Application}

In this section, we will show that non-equilibrium systems far from equilibrium can be described in a unified manner by Nambu non-equilibrium thermodynamics, with the BZ reaction representing time oscillations as a typical example of temporal dissipative structure, and the H-R model,
a model of spike-burst behavior of membrane potential, as Nambu non-equilibrium thermodynamics perspective.
\subsection{BZ reaction}

The BZ reaction is a typical case of a time-oscillating chemical reaction. 

Its simplest model is called Oregonator \cite{key-4}.

\begin{equation}
A+Y\stackrel{k_{1}}{\longrightarrow}X+P
\end{equation}

\begin{equation}
X+Y\stackrel{k_{2}}{\longrightarrow}2P
\end{equation}

\begin{equation}
A+X\stackrel{k_{3}}{\longrightarrow}2X+2Z
\end{equation}

\begin{equation}
2X\stackrel{k_{4}}{\longrightarrow}A+P
\end{equation}

\begin{equation}
B+Z\stackrel{k_{5}}{\longrightarrow}hY
\end{equation}
where $A=\mathrm{BrO_{3}^{-}},\ X=\mathrm{Br}\mathrm{O}_{2},\ Y=\mathrm{Br}^{-},\ Z=\mathrm{Ce^{4+}},\ P=\mathrm{H}\mathrm{Br}\mathrm{O}$,$\ B=\mathrm{\mathrm{CH_{2}(COOH)_{2}}}$
, $k_{1},\dots,k_{5}$ are reaction rate, and $h$ is an adjustable
stoichiometric factor. 

Normally, the concentrations $A$, $B$ and $P$ are assumed to be
constant.

Then, the equations of the time evolution of the BZ model become the
equations of motion for three coordinates or three concentrations
($X$, $Y$ and $Z$), given by

\begin{equation}
\frac{dX}{dt}=k_{1}AY-k_{2}XY+k_{3}AX-2k_{4}X^{2},
\end{equation}

\begin{equation}
\frac{dY}{dt}=-k_{1}AY-k_{2}XY+hk_{5}BZ,
\end{equation}

\begin{equation}
\frac{dZ}{dt}=2k_{3}AX-k_{5}BZ.
\end{equation}

These chemical reactions are far from equilibrium and are not obtained
from the Onsager's variational principle.
\color{black}
Following the methodology of Section 5, consider $Z$ as the catalyst, $H_{2}$.

We obtain

\begin{equation}
d\{H_{1},Z,x^{i}\}\wedge dx^{i}=d\dot{x}^i\wedge dx^{i}
\end{equation}

If we insert the concrete form from the development equation for velocity, $H_{1}$ must satisfy the following relation:

\begin{equation}
\begin{aligned}
&\frac{\partial^{2}H_{1}}{\partial X^{2}}+\frac{\partial^{2}H_{1}}{\partial Y^{2}}=k_{1}A-k_{2}(X-Y),\\
&\frac{\partial^{2}H_{1}}{\partial Z\partial X}=hk_{5}B, \frac{\partial^{2}H_{1}}{\partial Z\partial Y}=2k_{3}A. 
\end{aligned}
\end{equation}

From this,

\begin{equation}
\begin{aligned}
&\frac{\partial H_{1}}{\partial X}=hk_{5}BZ+f(X,Y), \\
&\frac{\partial H_{1}}{\partial Y}=2k_{3}AZ+g(X,Y), \\
&\frac{\partial f(X,Y)}{\partial X}+\frac{\partial g(X,Y)}{\partial X}=k_{1}X-k_{2}(X-Y).
\end{aligned}
\end{equation}

By substituting these into the above formula, 

\begin{equation}
f(X,Y)=\frac{1}{2}k_{2}X^{2},
\end{equation}

\begin{equation}
g(X,Y)=-k_{1}AY+\frac{k_{2}}{2}Y^{2}.
\end{equation}

Therefore, $H_{1}$ is obtained as 

\begin{equation}
H_{1}=hk_{5}BZX-\frac{1}{2}k_{1}AY^{2}+\frac{k_{2}}{6}Y^{3}+2k_{3}AYZ+\frac{1}{6}k_{2}X^{3}.
\end{equation}

The time evolution of the non-Hamiltonian part is necessarily in the form of an entropy gradient from the discussion in Section 5.



\begin{equation}
\begin{aligned}S= & -\frac{k_{2}}{2}XY^{2}-\frac{k_{2}}{2}X^{2}Y\\
 & -\frac{2}{3}k_{4}X^{3}+\frac{1}{2}k_{3}AX^{2}-\frac{1}{2}k_{1}AY^{2}\\
 & -\frac{k_{5}}{2}BZ^{2}+2k_{3}AXZ.
\end{aligned}
\end{equation}
\color{black}
We divide time evolution into Hamiltonian part and entropy part: 

\begin{equation}
\frac{d}{dt}=\partial_{t}^{(H)}+\partial_{t}^{(S)}.
\end{equation}

Then, we obtain the time evolution induced by the Hamiltonians, as
follows:
\begin{equation}
\partial_{t}^{(H)}X=-\{H_{1},H_{2},X\}=k_{1}AY+\frac{k_{2}}{2}Y^{2}-2k_{3}AZ,
\end{equation}

\begin{equation}
\partial_{t}^{(H)}Y=-\{H_{1},H_{2},Y\}=\frac{k_{2}}{2}X^{2}+hk_{5}BZ,
\end{equation}

\begin{equation}
\partial_{t}^{(H)}Z=-\{H_{1},H_{2},Z\}=0,
\end{equation}
while the time evolution induced by the entropy is given by

\begin{equation}
\partial_{t}^{(S)}X=\{S,Y,Z\}=k_{3}AX-2k_{4}X^{2}-k_{2}XY-\frac{k_{2}}{2}Y^{2}+2k_{3}AZ,
\end{equation}

\begin{equation}
\partial_{t}^{(S)}Y=\{S,Z,X\}=-\frac{k_{2}}{2}X^{2}-k_{1}AY-k_{2}XY,
\end{equation}

\begin{equation}
\partial_{t}^{(S)}Z=\{S,X,Y\}=2k_{3}AX-k_{5}BZ.
\end{equation}

Combining these two kinds of time evolutions gives the time evolution
of the BZ reaction.

Next, let us examine the time evolution of entropy.

It is written as follows:

\begin{equation}
\dot{S}=\partial_{t}^{(H)}S+\partial_{t}^{(S)}S,
\end{equation}

\begin{equation}
\partial_{t}^{(H)}S=-\{H_{1},H_{2},S\},
\end{equation}

\begin{equation}
\partial_{t}^{(S)}S=+\{S,X,Y\}^{2}+\{S,Y,Z\}^{2}+\{S,Z,X\}^{2}
\end{equation}
where

\begin{equation}
\begin{aligned}-\{H_{1},H_{2},S\} & =-\left(k_{1}AY+\frac{k_{2}Y^{2}}{2}-2k_{3}AZ\right)\\
\times & \left(-k_{3}AX+2k_{4}X^{2}+k_{2}XY+\frac{k_{2}Y^{2}}{2}-2k_{3}AZ\right)\\
 & -\left(\frac{k_{2}X^{2}}{2}+k_{1}AY+k_{2}XY\right)\\
 & \times\left(\frac{k_{2}X^{2}}{2}-k_{5}hBZ\right)
\end{aligned}
\end{equation}

\begin{equation}
=\color{black}-\color{black}\partial_{t}^{(H)}X\partial_{t}^{(S)}X\color{black}-\color{black}\partial_{t}^{(H)}Y\partial_{t}^{(S)}Y\color{black}-\color{black}\partial_{t}^{(H)}Z\partial_{t}^{(S)}Z.
\end{equation}

Thus we obtain
\begin{equation}
\dot{S}=-\partial_{t}^{(H)}X\partial_{t}^{(S)}X-\partial_{t}^{(H)}Y\partial_{t}^{(S)}Y-\partial_{t}^{(H)}Z\partial_{t}^{(S)}Z
\end{equation}

\begin{equation}
+\left(\partial_{t}^{(S)}X\right)^{2}+\left(\partial_{t}^{(S)}Y\right)^{2}+\left(\partial_{t}^{(S)}Z\right)^{2},
\end{equation}

\begin{equation}
\dot{H}_{1}=\partial_{t}^{(H)}H_{1}+\partial_{t}^{(S)}H_{1},
\end{equation}

\begin{equation}
\partial_{t}^{(H)}H_{1}=0,
\end{equation}

\begin{equation}
\partial_{t}^{(S)}H_{1}=\{H_{1},X,Y\}\{S,X,Y\}+\{H_{1},Y,Z\}\{S,Y,Z\}+\{H_{1},Z,X\}\{Z,X,S\}
\end{equation}

\begin{equation}
=\{H_{1},X,Y\}\partial_{t}^{(S)}Z+\{H_{1},Y,Z\}\partial_{t}^{(S)}X+\{H_{1},Z,X\}\partial_{t}^{(S)}Y,
\end{equation}

\begin{equation}
\{H_{1},X,Y\}=-k_{5}hBX+2k_{3}AY,
\end{equation}

\begin{equation}
\{H_{1},Y,Z\}=\frac{k_{2}X^{2}}{2}-k_{5}hBZ,
\end{equation}

\begin{equation}
\{H_{1},Z,X\}=-k_{1}AY-\frac{k_{2}Y^{2}}{2}+2k_{3}AZ,
\end{equation}

\begin{equation}
\dot{H}_{2}=\partial_{t}^{(H)}H_{2}+\partial_{t}^{(S)}H_{2},
\end{equation}

\begin{equation}
\partial_{t}^{(H)}H_{2}=0,
\end{equation}

\begin{equation}
\partial_{t}^{(S)}H_{2}=\partial_{t}^{(S)}Z=2k_{3}AX-k_{5}BZ.
\end{equation}

\begin{figure}[H]
\centering
\includegraphics[width=0.5\textwidth,height=0.3\linewidth]{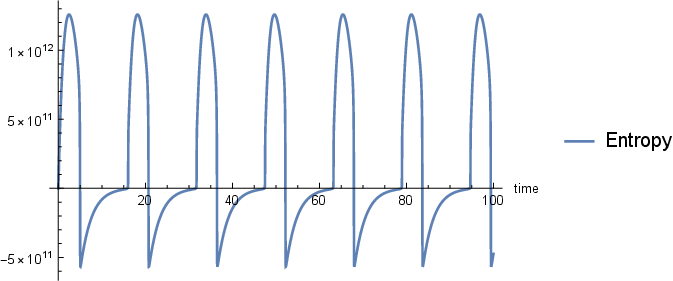}

\caption{Prot of entropy change over time. The horizontal axis represents time and the vertical axis represents entropy.\label{fig:Entropy}}
\end{figure}

\begin{figure}[H]
    \begin{tabular}{cc}
      \begin{minipage}[t]{0.45\hsize}
        \centering
        \includegraphics[width=1.0\textwidth,height=1.0\linewidth]{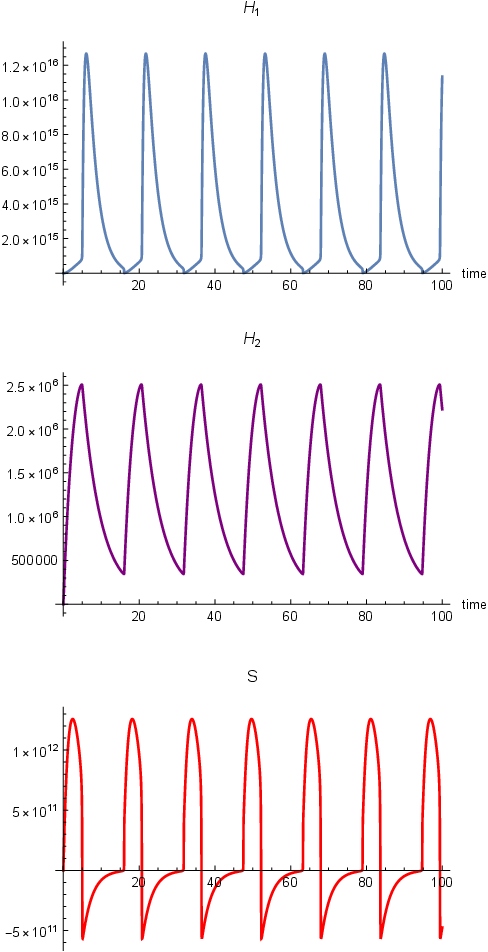}
        \caption{Plot of $H_1$, $H_2$ and $S$ as a function of time. The horizontal axis is time and the vertical axis is the values of $H_1$, $H_2$ and $S$.}
        \label{fig:H1H2S}
      \end{minipage} &
      \begin{minipage}[t]{0.45\hsize}
        \centering
        \includegraphics[width=1.0\textwidth,height=1.0\linewidth]{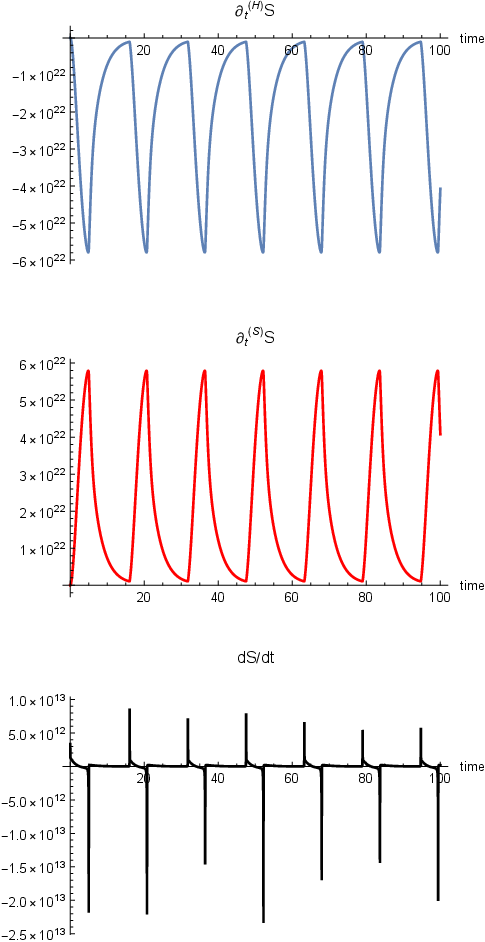}
        \caption{Plot of entropy velocity as a function of time. $\partial^{(H)}_t S$ and $\partial^{(S)}_t S$ represent the contribution of the Hamiltonian in the entropy rate and the dissipation due to entropy in the time variation of entropy, respectively, and $dS/dt$ is the sum of these contributions.}
        \label{fig:entropyVelocity}
      \end{minipage}
    \end{tabular}
  \end{figure}

\begin{figure}[H]
      \begin{tabular}{cc}
        \begin{minipage}[t]{0.45\hsize}
          \centering
          \includegraphics[width=1.0\textwidth,height=1.0\linewidth]{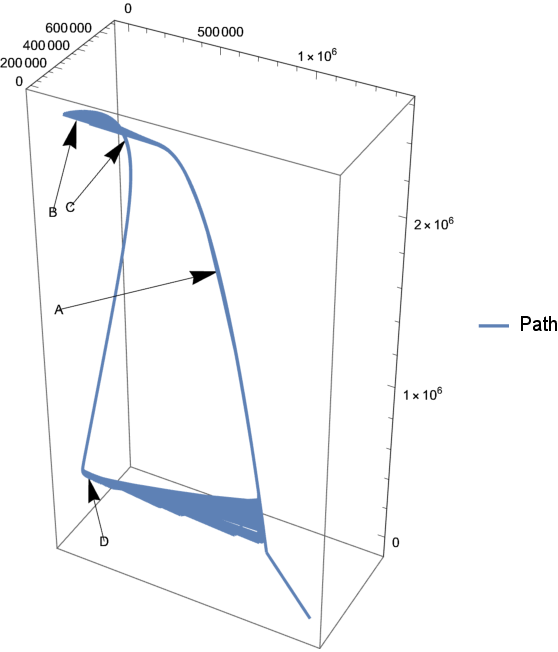}
        \end{minipage} &
        \begin{minipage}[t]{0.45\hsize}
          \centering
          \includegraphics[width=1.0\textwidth,height=1.0\linewidth]{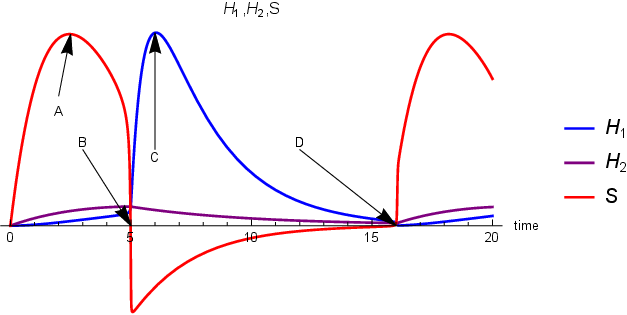}
        \end{minipage}
      \end{tabular}
      \caption{3D diagram of the limit cycle according to the time variation of the concentration X, Y and Z. The vertical axis represents the concentration of Z, the horizontal axis represents the concentration of X and the remaining axis represents the concentration of Y. \label{fig:Path}}
    
    \end{figure}

In this paper the time evolution of the BZ reaction is numerically
studied, which yields the following results depicted in Figures \ref{fig:X,Y,Z},
 \ref{fig:Entropy}, \ref{fig:H1H2S}, \ref{fig:entropyVelocity} and \ref{fig:Path}. These
figures illustrate how the BZ reaction oscillates.

Figure \ref{fig:X,Y,Z} shows the temporal development of the
concentrations, $X$, $Y$ and $Z$, as a function of time given in
the horizontal axis. Figure \ref{fig:Entropy} gives the time development of the
entropy $S$.

Figure \ref{fig:H1H2S} gives the change of the Hamiltonian $H_{1},H_{2}$
and the entropy $S$ in time.

Figure \ref{fig:entropyVelocity} shows the temporal change of entropy
$\frac{\partial S}{\partial t}$, $\frac{\partial^{(H)}S}{\partial t}$
and $\frac{\partial^{(S)}S}{\partial t}$ in detail.

As can be seen from Figure \ref{fig:entropyVelocity}, the respective
contributions from $\frac{\partial^{(H)}S}{\partial t}$ and $\frac{\partial^{(S)}S}{\partial t}$
almost cancel each other, and entropy is found to be almost unchanged,
except for the sudden increase and decrease of the entropy, subjected
to periodic delta-function-type positive and negative kicks. The positive
and negative kicks arise alternately, where the period of positive
kick $T_{+}$ and negative kick $T_{-}$ are both $T_{+}=T_{-}\approx20$.

This can be seen also in the path diagram, Figure\ref{fig:Path},
where the orbits undergo an inverted kick at both ends of the elongated
circle, and only then does entropy undergo an abrupt change. It can be seen from the figure that $H_1$ and $S$ are both zero at such a turning point.

Now, we can understand well why the usual non-equilibrium thermodynamics
by Onsager can not be applied to the BZ reaction. The Onsager theory
is formulated near the equilibrium point of the entropy. 
There, the entropy gradient describes the time evolution of the thermodynamic variable as an affinity force, 
and as time passes, entropy increases while its velocity becomes zero.
In systems far from equilibrium,  dynamics are described by entropy gradients plus Nambu dynamics.
In systems such as the BZ reaction, entropy does not necessarily increase, but rather oscillates between the Hamiltonians $H_1$ and $H_2$.

\begin{figure}[H]
  \includegraphics[width=1\textwidth,height=0.45\textwidth]{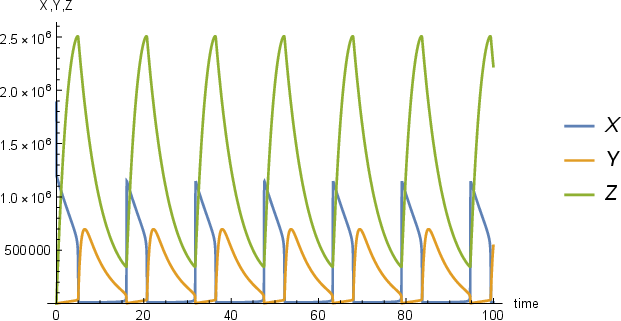}
  \begin{center}
  \caption{Time variation of concentration X, Y and Z. Horizontal axis represents time, vertical axis represents concentration.\label{fig:X,Y,Z}}
  \par\end{center}%
\end{figure}

\color{black}

    Note that oscillations do not occur in all parameter regions. In particular, in the case of $ hk_{5}\sim k_{1}A\sim k_{2}\sim k_{4}X\sim k_{5}B\ll k_{3}A$, the BZ reaction is an oscillatory phenomenon. Approximating the Hamiltonian and entropy in the case, we get 
    \begin{equation}
      \begin{aligned}
        &H_{1}\approx2k_{3}AYZ+\dots \\
        &H_{2}=Z, \\
        &S\approx\frac{1}{2}k_{3}AX^{2}+2k_{3}AXZ+\dots.
      \end{aligned}
    \end{equation}
    
    This represents a system of harmonic oscillators as discussed in Appendix E. Thus, under the appropriate variable transformations, the system is approximately described as a system of harmonic oscillator Hamiltonians with no entropy.
    Thus, from the given Hamiltonian and entropy, we can analyze whether oscillatory phenomena occur    \footnote{When the condition \(*dS = dH_1 \wedge d\Delta H_2 + d\Delta H_1\wedge dH_2 + d\Delta H_1\wedge d\Delta H_2\) is satisfied, the redefinition \(H_i \rightarrow H_i + \Delta H_i\) leads to the elimination of the entropy \(S\), leaving only the Hamiltonians to describe the system. This removal of \(S\) is crucial in achieving a quasi-conserved system of Hamiltonians. In general, the interaction between the set of Hamiltonians and entropy plays a significant role in determining the system's behavior.}
    .

\color{black}

\subsection{Hindmarsh-Rose model}
The H-R model is a biological neuron model of the spike-burst behavior of the membrane potential which is a typical behavior of neurons in the brain\cite{key-51}\cite{key-52}. The H-R model is a three-dimensional dynamical system, 
which is given by the following equations:
\begin{align}
&\dot{x}=y+\phi(x)-z+I, \\
&\dot{y}=\psi(x) - y, \\
&\dot{z}=r(s(x - x_R)-z),
\end{align}
where
\begin{align}
  &\phi(x) = -a x^3 + b x^2, \\
  &\psi(x) = c - d x^2, 
\end{align}
where $x$ is the biological membrane potential and $x_R$ is the resting potential.
$y$ is called the recovery variable as it takes into account the transport of ions across the membrane through the ion channel.
$z$ is the bursting variable and $I$ is the current entering the neuron from outside and is used as a control parameter.
$a,b,c,d,s$ are fixed parameters of the model and often take values such as $a=1, b=3, c=1, d=5, s=4$.
Also, $r$ is associated with timescales related to neural adaptation and takes on a very small order, such as $10^{-3}$.

\color{black}
From the same discussion as for the BZ reaction, we obtain

\begin{equation}
  \begin{aligned}
& \frac{\partial^{2}H_{1}}{\partial X^{2}}+\frac{\partial^{2}H_{1}}{\partial Y^{2}}=1+2dX, \\
& -\frac{\partial^{2}H_{1}}{\partial Z\partial X}=0, \\
& \frac{\partial^{2}H_{1}}{\partial Z\partial Y}=1+rs. \\
  \end{aligned}
\end{equation}

From this,

\begin{equation}
  \begin{aligned}
&\frac{\partial H_{1}}{\partial X}=+f(X,Y),\\
&\frac{\partial H_{1}}{\partial Y}=(1+rs)Z+g(X,Y), \\
&\frac{\partial f(X,Y)}{\partial X}+\frac{\partial g(X,Y)}{\partial X}=1+2dX. \\
\end{aligned}
\end{equation}

By substituting these into the above formula, 

\begin{equation}
f(X,Y)=-dX^{2},
\end{equation}
\begin{equation}
g(X,Y)=-Y.
\end{equation}

Therefore, $H_{1}$ is obtained as 

\begin{equation}
H_{1}=-\frac{d}{3}X^{3}-\frac{1}{2}Y^{2}+(1+rs)YZ.
\end{equation}

In the Nambu non-equilibrium thermodynamics, $H_1$, $H_2$, and $S$ corresponding to this model have the following forms:
\begin{align}
  &H_1 = -\frac{d}{3} x^3 - \frac{1}{2}y^2 + (1 + r s) y z, \\
  &H_2 = z, \\
  &S = - \frac{a}{4} x^4 + \frac{b}{3} x^3 + I x - \frac{y^2}{2} + c y - \frac{r z^2}{2} - r s x_R z + r s x z.
\end{align}
\color{black}

Also, we divide time evolution into Hamiltonian part and entropy part as follows:

\begin{align}
  &\partial_{t}^{(H)}x=-\{H_{1},H_{2},x\}= y - (1+r s) z, \\
  &\partial_{t}^{(H)}y=-\{H_{1},H_{2},y\}= - d x^2, \\
  &\partial_{t}^{(H)}z=-\{H_{1},H_{2},z\}= 0,
\end{align}

\begin{align}
  &\partial_{t}^{(S)}x= I + b x^2 - a x^3 + r s z, \\
  &\partial_{t}^{(S)}y= c - y \\
  &\partial_{t}^{(S)}z= r(s(x - x_R)-z).
\end{align}

Next, the time evolution of entropy is given by

\begin{equation}
  \dot{S} = \partial_{t}^{(H)}S + \partial_{t}^{(S)}S, \\
\end{equation}
where
\begin{align}
  &\partial^{(H)}_t S  = -d x^2 (c - y) - (I + b x^2 - a x^3 + r s z) (-y + (1 + r s) z), \\
  &\partial^{(S)}_t S  =  (c - y)^2 + r^2(s(x_R-x) + z)^2 + (I + (b - a x)x^2 + r s z)^2.
\end{align}

the time evolution of Hamiltonian is given by
\begin{align}
  \dot{H_1} = &- r (1 + r s) y (s (x - x_R) -z) + d x^2 ( I + x^2 (b  - a x) + rsz)  \\
  &+ (c-y) (y - (1 + r s) z), \nonumber\\
  \dot{H_2} = &r(s(x_R - x)+z).
\end{align}

\begin{figure}[H]
\includegraphics[width=1\textwidth,height=0.45\textwidth]{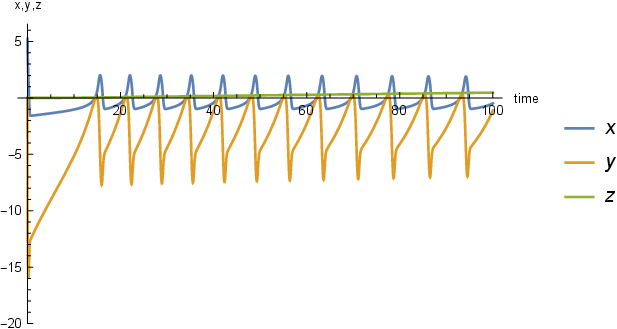}
\begin{center}
\caption{Time variation of membrane potential $x$, recovery variable $y$ and bursting variable $z$. Horizontal axis represents time, vertical axis represents $x$, $y$, $z$.\label{fig:rosexyz}}

\par\end{center}%
\end{figure}

\begin{figure}[H]
\includegraphics[width=0.5\textheight,height=0.5\textwidth]{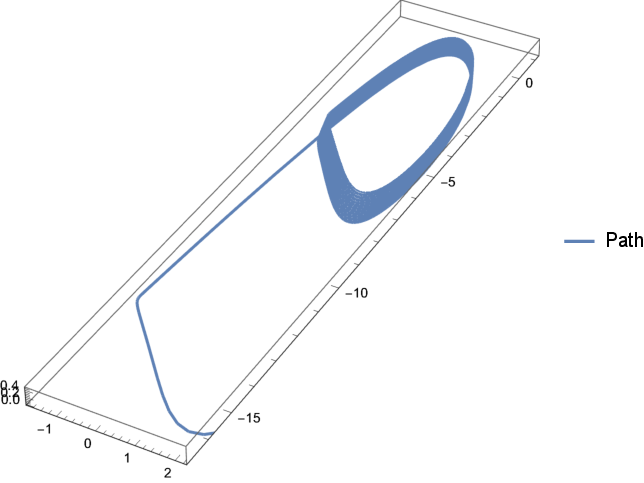}
\begin{center}
\caption{3D diagram of the limit cycle according to the time variation of $x$, $y$ and $z$. The vertical axis represents the bursting variable $z$, the horizontal axis represents the membrane potential $x$ and the remaining axis represents the recovery variable $y$. \label{fig:rosePath}}

\par\end{center}%
\end{figure}

\begin{figure}[H]
  \begin{tabular}{cc}
    \begin{minipage}[t]{0.45\hsize}
      \centering
      \includegraphics[width=1.0\textwidth,height=1.0\linewidth]{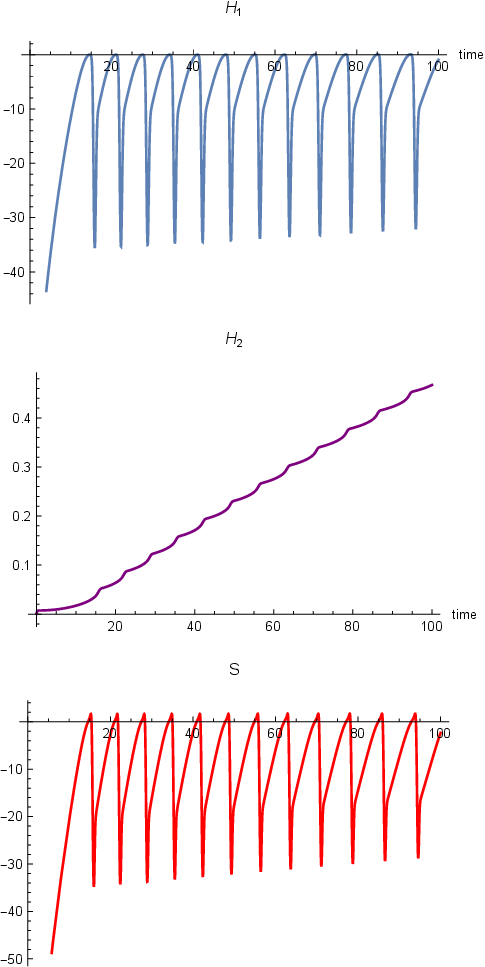}
      \caption{Plot of $H_1$, $H_2$ and $S$ as a function of time. The horizontal axis is time and the vertical axis is the values of $H_1$, $H_2$ and $S$.}
      \label{fig:roseH1H2S}
    \end{minipage} &
    \begin{minipage}[t]{0.45\hsize}
      \centering
      \includegraphics[width=1.0\textwidth,height=1.0\linewidth]{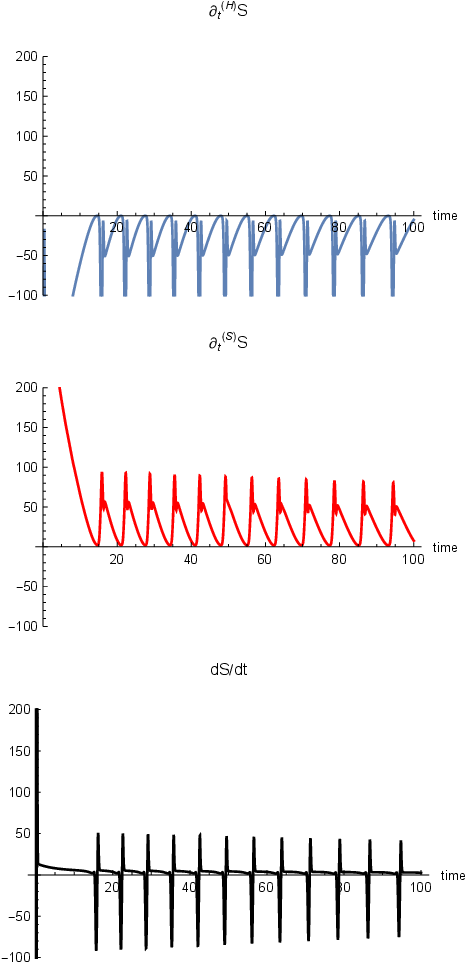}
      \caption{Plot of entropy velocity as a function of time. $\partial^{(H)}_t S$ and $\partial^{(S)}_t S$ represent the contribution of the Hamiltonian in the entropy rate and the dissipation due to entropy in the time variation of entropy, respectively, and $dS/dt$ is the sum of these contributions.}
      \label{fig:roseEntropyVelocity}
    \end{minipage}
  \end{tabular}
\end{figure}

The spike-like excitatory activity of the membrane potential is shown in Figure \ref{fig:rosexyz} and Figure \ref{fig:rosePath}. The trajectory shows a limit cycle.

Figure \ref{fig:roseH1H2S}  shows the time evolution of $H_1$, $H_2$, and $S$. It can be seen that $H_1$ and $S$ take alternating peaks in a spike-like pattern, while $H_2$ increases in a staircase-like pattern.

Figure \ref{fig:roseEntropyVelocity} decomposes the time evolution of the rate of entropy into a contribution from the Hamiltonian and a contribution from entropy, with the Hamiltonian contribution of entropy having a periodic contribution to entropy.

\section{Analysis of cycles and spikes}
\begin{figure}[H]
    \begin{tabular}{cc}
      \begin{minipage}[t]{0.45\hsize}
        \centering
        \includegraphics[width=1.0\textwidth,height=1.0\linewidth]{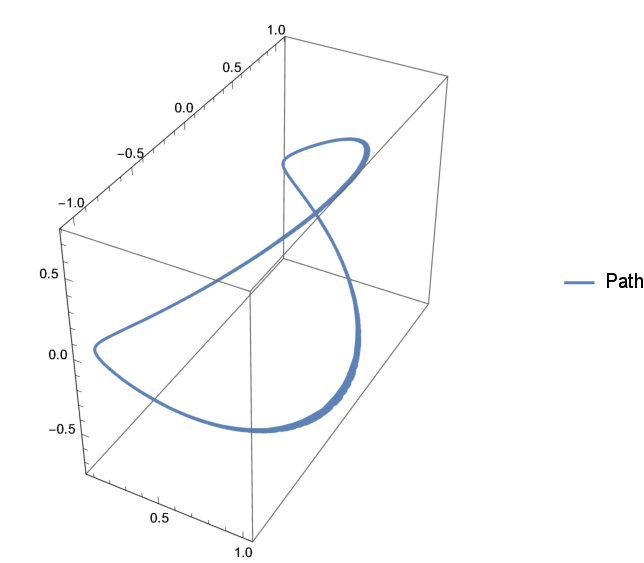}
        \caption{3D diagram of the limit cycle according to the time variation of $x$, $y$ and $z$.}
        \label{fig:spikepath}
      \end{minipage} &
      \begin{minipage}[t]{0.45\hsize}
        \centering
        \includegraphics[width=1.0\textwidth,height=1.0\linewidth]{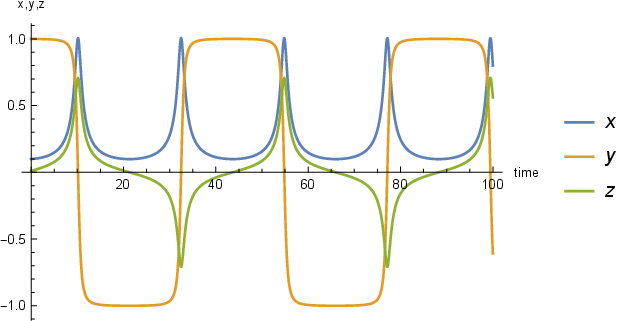}
        \caption{Plot of $x$, $y$ and $z$ as a function of time.}
        \label{fig:spikexyz}
      \end{minipage}
    \end{tabular}
  \end{figure}

Two characteristic behaviors of systems far from equilibrium are the cycle of trajectories and the presence of spikes.
If the system has conserved quantities, it is expected that the trajectories will cycle because they must always keep moving to satisfy the conserved quantities.
 Even if there are no conserved quantities, a cycle can exist for a constant long time if there are quantities that can be regarded as pseudo-conserved quantities such that they are asymptotically decreasing.
The triangular reaction was already an example that has such a conserved quantity, but we can also consider the following as a simple toy model.
\begin{equation}
H_{1}=a-x^{3}z,
\end{equation}
\begin{equation}
H_{2}=b+z,
\end{equation}
\begin{equation}
S=x^{2}yz.
\end{equation}
The equations of motion are as follows
\begin{equation}
\dot{x}=2xyz,
\end{equation}
\begin{equation}
\dot{y}=-2x^{2}z,
\end{equation}
\begin{equation}
\dot{z}=x^{2}y.
\end{equation}
This preserves 
\begin{equation}
O=\frac{1}{2}(x^{2}+y^{2})
\end{equation}
in time evolution.
The path draws a limit cycle as in Figures \ref{fig:spikepath} and \ref{fig:spikexyz}, with a sharp spike for the $x$ and $z$ components resulting from its nonlinearity.

Another toy model is 
\begin{equation}
    H_1 = H_1(x,y,z),
\end{equation}
\begin{equation}
    H_2 = (x^2 - y^2) f(z),
\end{equation}
\begin{equation}
    S = xy,
\end{equation}
where $H_1(x,y,z)$ and $f(z)$ are arbitrary functions. The equations of motion are
\begin{equation}
\dot{x}=y-2yf(z)\frac{\partial H_{1}}{\partial z}+(y^{2}-x^{2})\frac{df(z)}{dz}\frac{\partial H_{1}}{\partial y},
\end{equation}
\begin{equation}
\dot{y}=x-2xf(z)\frac{\partial H_{1}}{\partial z}-(y^{2}-x^{2})\frac{df(z)}{dz}\frac{\partial H_{1}}{\partial x},
\end{equation}
\begin{equation}
\dot{z}=2xf(z)\frac{\partial H_{1}}{\partial y}+2yf(z)\frac{\partial H_{1}}{\partial x}.
\end{equation}

In this model, $H_2$ itself is the conserved quantity of time evolution.

As a concrete example, we take
\begin{equation}
    H_1 = \frac{1}{2}(x^2 + y^2 + z^2),
\end{equation}
\begin{equation}
    f(z) = z^2.
\end{equation}

The equations of motion are
\begin{equation}
\dot{x}=y-2x^{2}yz+2y^{3}z-2yz^{3},
\label{eq:spikeToyModel}
\end{equation}
\begin{equation}
\dot{y}=x+2x^{3}z-2xy^{2}z-2xz^{3},
\label{eq:spikeToyModel2}
\end{equation}
\begin{equation}
\dot{z}=4xyz^{2}.
\label{eq:spikeToyModel3}
\end{equation}

\begin{figure}[H]
    \begin{tabular}{cc}
      \begin{minipage}[t]{0.45\hsize}
        \centering
        \includegraphics[width=1.0\textwidth,height=1.0\linewidth]{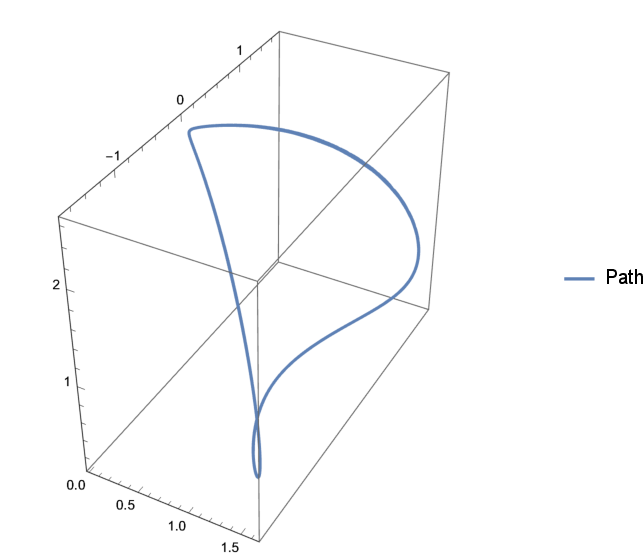}
        \caption{3D diagram of the limit cycle according to the time variation of $x$, $y$ and $z$.}
        \label{fig:h1spikepath}
      \end{minipage} &
      \begin{minipage}[t]{0.45\hsize}
        \centering
        \includegraphics[width=1.0\textwidth,height=1.0\linewidth]{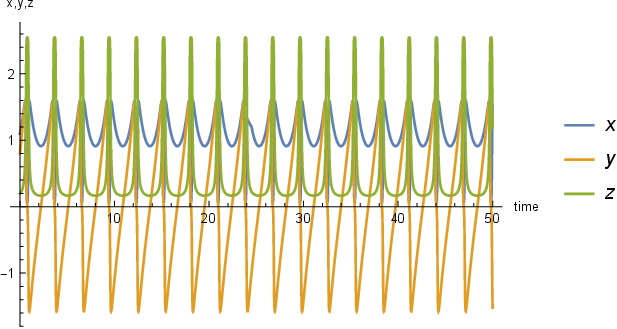}
        \caption{Plot of $x$, $y$ and $z$ as a function of time.}
        \label{fig:h1spikexyz}
      \end{minipage}
    \end{tabular}
  \end{figure}

The path draws a limit cycle as in Figures \ref{fig:h1spikepath} and \ref{fig:h1spikexyz}, with a sharp spike for the $y$ and $z$ components resulting from its nonlinearity.

\section{Fluctuation effects in Nambu non-equilibrium thermodynamics}

\color{black}
In previous sections, we simply added the effect of thermal fluctuations to the time evolution.
In this section, let discuss how this thermal fluctuation affects the results of numerical calculations in a concrete system.\footnote{\color{black}
The Fokker-Planck equation describing nonequilibrium thermodynamics is an equation similar to the Sch{\"o}dinger equation in quantum mechanics. Thus, nonequilibrium statistical mechanics (stochastic processes) can also be described by a ``noncommutative operator pair'' of a thermodynamic variable and its differential operator. Then, the fluctuation term is related to the problem of quantization of the Nambu bracket and is of interest in discussing the quantization of membrane  in high
energy theory (2-brane)\cite{key-8}\color{black}.}. 
Therefore, in BZ reaction, we introduce a time evolution equation with fluctuation effects as follows,

\color{black}
\begin{equation}
\frac{dX}{dt}=k_{1}AY-k_{2}XY+k_{3}AX-2k_{4}X^{2}  + \kappa  \xi_1,
\end{equation}

\begin{equation}
\frac{dY}{dt}=-k_{1}AY-k_{2}XY+hk_{5}BZ + \kappa \xi_2,
\end{equation}

\begin{equation}
\frac{dZ}{dt}=2k_{3}AX-k_{5}BZ + \kappa \xi_3.
\end{equation}

\begin{figure}[t]
    \begin{tabular}{cc}
      \begin{minipage}[t]{0.45\hsize}
        \centering
        \includegraphics[width=1.0\textwidth,height=1.0\linewidth]{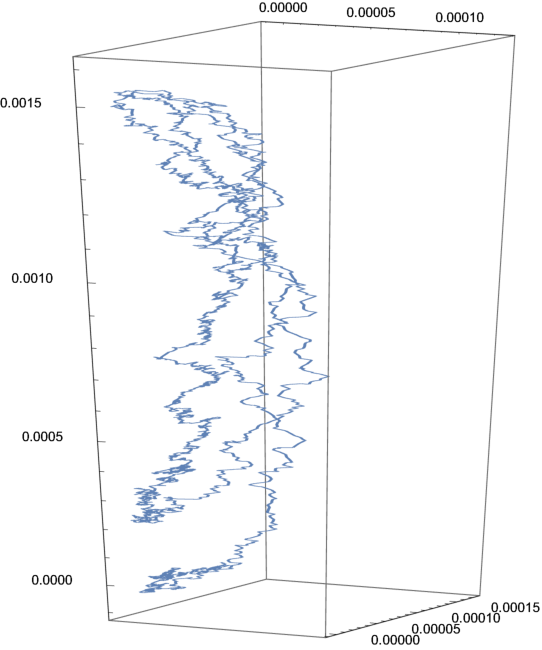}
        \caption{Path with fluctuation.}
        \label{fig:case1bzfl}
      \end{minipage} &
      \begin{minipage}[t]{0.45\hsize}
        \centering
        \includegraphics[width=1.0\textwidth,height=1.0\linewidth]{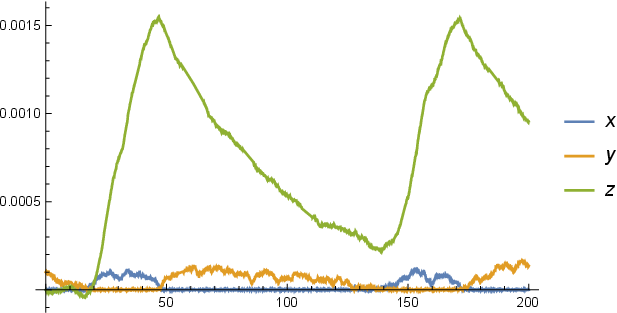}
        \caption{$X$, $Y$ and $Z$ with fluctuation.}
        \label{fig:case1bzflxyz}
      \end{minipage}
  \end{tabular}
  \begin{tabular}{cc}
    \begin{minipage}[t]{0.45\hsize}
      \centering
      \includegraphics[width=1.0\textwidth,height=1.0\linewidth]{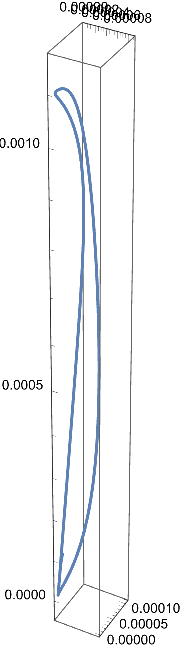}
      \caption{Path with no fluctuation.}
      \label{fig:case1bznofl}
    \end{minipage} &
    \begin{minipage}[t]{0.45\hsize}
      \centering
      \includegraphics[width=1.0\textwidth,height=1.0\linewidth]{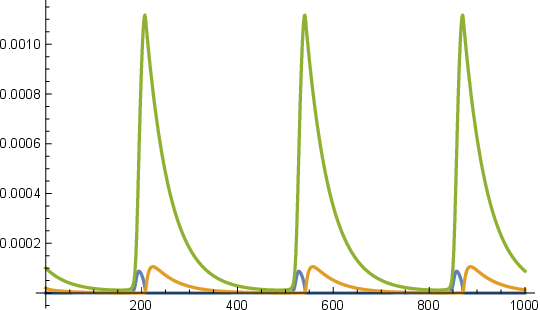}
      \caption{$X$, $Y$ and $Z$ with no fluctuation.}
      \label{fig:case1bznoflxyz}
    \end{minipage}
\end{tabular}
\end{figure}

\color{black}
The numerical results for each case are presented in Figures \ref{fig:case1bzfl} and \ref{fig:case1bzflxyz}.
Figure \ref{fig:case1bznofl} and \ref{fig:case1bznoflxyz}  also shows the path in the
case of no fluctuation for comparison.\color{black}

The parameters used are as follows: $k_{1}=1.28,\ k_{2}=8.0,\ k_{3}=8.0\times10^5,\ k_{4}=2.0\times 10^3,\ k_{5}=1.0,\ h=1.5,\ A=0.06,\ B=0.02,\ X(0)=2\times 10^{-7},Y(0)=0.0001,\ Z(0)=0.0, \kappa=10^{-5}$
and $dt=10^{-2}$ for split time interval.

As can be seen by comparison of the figures, the addition of white
noise simply causes the pathway to fluctuate slightly around the noiseless
case.
It can be seen that the limit cycle does not lose its time periodicity, although its trajectory changes with each cycle by adding fluctuations.
\color{black}

\section{Summary and discussion}

We have shown that the use of Nambu brackets is effective in describing
non-equilibrium thermodynamics in situations far from equilibrium,
taking a chemical reaction (the BZ reaction) and H-R model as an example.

We have found that a reaction model, such as the BZ reaction, must
include an entropic contribution in addition to multiple Hamiltonians. 
This differs from the previously known Lotka-Volterra model, which
can be described by only Nambu brackets and only two Hamiltonians.

The Lotka-Volterra model requires no entropy when there are three
free variables and the two Hamiltonians are conserved quantities.
On the other hand, in the BZ model we are discussing, the two Hamiltonians
are not conserved quantities. 

As is examined in this paper, the time evolution of entropy in the
time-oscillating BZ reaction shows that the Hamiltonian and entropy
contributions cancel with each other out, and hence the entropy does
not increase much except at the end points of the cycle. 

\color{black}

In recent decades, there have been several attempts to extend Onsager's near-equilibrium thermodynamics to far-from-equilibrium systems, including the well-known Glansdorff-Prigogine theory\cite{Prigogine}.
This approach is, however, not so powerful to describe oscillatory behaviors or to grasp the general features of far-from-equilibrium systems.
Glansdorff-Prigogine's universal evolution criterion can be applied not only to the linear region near equilibrium but also to certain non-linear regions. It provides a way to analyze the time evolution of entropy production in systems far from equilibrium\cite{Prigogine}. 
However, while it is useful in these contexts, its application is limited when dealing with more complex, oscillatory phenomena that involve multiple dynamical variables or Hamiltonians. In such cases, additional theoretical frameworks are required.
In contrast, the Nambu non-equilibrium thermodynamics introduced in this paper allows for the analysis of non-linear and oscillatory phenomena through the inclusion of Hamiltonian parts, providing a new framework that extends beyond traditional thermodynamic theories.

Additionally, Ziegler's maximum entropy production principle provides an important perspective on non-equilibrium thermodynamics, particularly in relation to systems with non-linear dynamics\cite{MEPP}.
This principle is closely related to Jaynes' maximum entropy approach in statistical mechanics\cite{Jaynes}, which maximizes the information entropy to predict the most likely distribution of states.  \color{black}Furthermore, Ziegler's principle has connections to the large deviation principle, which describes the probability of rare fluctuations far from equilibrium\cite{LargeFluctuations}.

Nambu non-equilibrium thermodynamics is built to cope with the phenomena far from the equilibrium, by introducing multi-Hamiltonians and a thermodynamic entropy, which allows to analyze the systems exhibiting both entropy-driven and Hamiltonian-driven behaviors, such as the BZ reaction and Hindmarsh-Rose model.  These oscillatory behaviors with cycles and spikes cannot be fully captured by Ziegler's or Jaynes' frameworks alone.
\color{black}

Numerical simulations have been carried out in detail with respect
to the temporal evolution of the chemical reactions with thermal fluctuations.

\color{black}
 In this paper we can put forward a general framework for the ``Nambu non-equilibrium thermodynamics''.  In this formalism, it is shown that any autonomous differential equations for $N$ variables can be written locally (or temporarily) as a system of Nambu non-equilibrium thermodynamics, having $(N-1)$ Hamiltonians and an entropy $S$.  \color{black}Since the global existence of multi-Hamltonians is not guaranteed rigorously, we assume the existence of them a priori.  It is important to note that this formalism is applicable to the systems far from equilibrium, having non-linear response and cycles.  \color{black}Then, we may expect ``another second law of thermodynamics'', which drives the highly non-equilibrium system to a special system $\Omega_{\text{cycle}}$, in which the cycle exists and the third entropy $S^C$ becomes a constant.

One advantage of using Nambu brackets is that they clarify the structure.

 We can analyze the motive forces of the time development separately, one by Hamiltonian, and the other by entropy.  The contribution of them cannot be  separated by the numerical calculation of the differential equation in the BZ reaction and others.  Nambu theory is a theory that describes multiple conservation laws with multiple Hamiltonians $H_1\dots H_n$. The introduction of entropy may break some of these conservation laws, but the remaining unbroken conservation laws or pseudo-conservation laws (periodicity) remain, which leads to the generation of cycles.

When cycle structures are formed in systems far from equilibrium in Nambu non-equilibrium thermodynamics, their entropy can be negative in time evolution associated with the Hamiltonian, while the time evolution associated with the entropy is always positive.   This oscillation of entropy is related to the formation of cycle, which may trigger the structure formation. \color{black} This provides some insight into the role of entropy in the formation of structure.

According to non-linear response theory, there is a close relationship
between fluctuations and the advection term.

In this paper, we add white noise to the effects of fluctuations through discussion in a microscopic model\color{black}, but we need
to examine which types of fluctuations are appropriate
to be included in the non-linear response theory in the Nambu non-equilibrium
thermodynamics. Also, although this paper discusses the case where
chemical reactions are location-independent, it is easy to extend
it to the location-dependent case and it would be interesting to investigate
how the fluctuation effects contribute to
the formation of structures such as Turing patterns and others. It
is also interesting to discuss other time-oscillating models such
as the BL model\cite{key-1} and the chaotic behavior of the Lorenz
model\cite{key-lorenz} from the perspective of Nambu non-equilibrium thermodynamics.

In addition, continued research is also needed to determine the generality of the system described by Nambu non-equilibrium thermodynamics.

Further studies on these points are expected in the near future.

\section*{Acknowledgments}

We would like to thank Toshio Fukumi about non-linear response theory.
We are indebted to Shiro Komata for reading this paper and giving
useful comments. 

\color{black}
\section*{Appendix A. Entropy in LCR circuits}
\renewcommand{\theequation}{A.\arabic{equation}}
\setcounter{equation}{0} 
In this section, we discuss that $S$ in the non-equilibrium thermodynamic treatment in LCR can be regarded as thermodynamic entropy.
In the LCR circuit, the time derivative of the total energy is
\begin{equation}
  \dot{E} = IV = RI^2 + L\dot{I}I + \frac{Q}{C} I.
\end{equation}
The thermal energy per unit time due to resistance is 
\begin{equation}
  \dot{Q_h} = RI^2.
\end{equation}
From the first law of thermodynamics, the time derivative of work is 
\begin{align}
  \dot{W} &= \dot{E} - \dot{Q}  = \frac{d}{dt}\left( \frac{1}{2} LI^2 + \frac{Q^2}{2C}\right) \\
&= \frac{d}{dt} \left( \frac{\Phi^2}{2L} + \frac{Q^2}{2C} \right) = \frac{d}{dt}H
\end{align}
From the entropy generation written by
\begin{equation}
  \dot{Q}_h = T\partial^{(S)}_tS = TL^{ij} \frac{\partial S}{\partial{x^i}}\frac{\partial S}{\partial{x^j}} = 2 \Psi,
\end{equation}
the entropy generation is
\begin{equation}
  T\partial^{(S)}_tS = RI^2 = R^{-1} \left( \frac{\partial}{\partial \Phi} \frac{R\Phi^2}{2L} \right)^2.
\end{equation}

Therefore, the entropy is written by
\begin{equation}
  S = \frac{R \Phi^2}{2L}.
\end{equation}

This is in agreement with the entropy introduced in Section 2, which shows that $S$ is the entropy in thermodynamics.

\section*{Appendix B. Turning points of spikes}
\renewcommand{\theequation}{B.\arabic{equation}}
\setcounter{equation}{0} 

In this section, we discuss the turning points of spikes in the toy model in Section 4.

From the equations of motion(\ref{eq:spikeToyModel})-(\ref{eq:spikeToyModel3}), we obtain the following equations,
\begin{align}
    dt &= \frac{dx}{y-2x^{2}yz+2y^{3}z-2yz^{3}} \\
       & = \frac{dy}{x+2x^{3}z-2xy^{2}z-2xz^{3}} \\
        & = \frac{dz}{4xyz^{2}}.
\end{align}
Now consider a cusp that changes abruptly only in the $x$ direction.
Since the denominator is quadratic for $x$, the equation can be rearranged as follows.
\begin{equation}
    dt = \frac{dx}{(x-\alpha_+)(x-\alpha_-)} = dx\left(\frac{A_+}{x-\alpha_+}+\frac{A_-}{x-\alpha_-}\right),
\end{equation}
\begin{equation}
    \alpha_\pm \equiv \mp \sqrt{y^2-z^2+\frac{1}{2 z}},
\end{equation}
\begin{equation}
    A_\pm = 2\alpha_\pm.
\end{equation}

Since $dt$ is singular in terms of the cusp arising and 
only the contribution of the $A_+$ term works around $x = \alpha_+$,
 we can write the following equation.

\begin{equation}
    \frac{x(t)-\alpha_+}{x(t_0) - \alpha_+} = e^{(t-t_0)/A_+} = (t-t_0)/A_+ + O((t-t_0)^2/A_+^2)
\end{equation}

Therefore, the gap before and after the turning point can be estimated from the following equation:
\begin{align}
    \dot{x}(t_* + \delta t) - \dot{x}(t_* - \delta t) &= 1/A_+ \left(x(t_* + \delta t) + x(t_* - \delta t)\right) \\
     & = \frac{1}{2\sqrt{y(t_*)^2-z(t_*)^2+\frac{1}{2 z(t_*)}}}\left(x(t_* + \delta t) + x(t_* - \delta t)\right), \\
\end{align}
where $t_*$ is time at the turning point.

\color{black}
\section*{Appendix C. Zwanzig's model}
\renewcommand{\theequation}{C.\arabic{equation}}
\setcounter{equation}{0} 

In this section, we discuss the derivation of fluctuation effects from microscopic models in an extended discussion of Zwanzig's model\cite{key-60}.

Consider the following system of slow variables $X_l = X, Y, Z$ coupled to a fast variable, i.e., harmonic oscillation $x_i$, which can be regarded as a heat bath:

\begin{align}
  &H_1 = H^0_1(\{X\}) + \sum^N_{i=1}\sum^3_{l=1} \frac{{p^i_l}^2}{2} + \sum^N_{i=1} \sum^3_{l=1}m {\omega^i_l}^2\left(x^i_l - \frac{\gamma^i_l}{2\omega^i_l}X_l\right), \\
  &H_2 = H^0_2(\{X\}) + \sum^N_{i=1}\sum^3_{l=1} z^i_l, \\
  &S = S(\{X\}),
\end{align}

Therefore, the time evolution for $x$, $y$, $z$ is 

\begin{align}
  &\dot{x}^i_l = \frac{p^i_l }{m}, \\
  &\dot{p}^i_l = m {\omega^i_l}^2 (x^i_l - \frac{\gamma^i_l}{2\omega^i_l}X_l), \\
  &\dot{z}^i_l = 0.
\end{align}
This differential equation can be solved and the solution is 
\begin{align}
  &x^i_l(t) - \frac{\gamma^i_l}{m {\omega^i_l}^2}X_l(t) = - \int^t_0 \frac{\gamma^i_l}{m 
  {\omega^i_l}^2}\frac{d X_l(s)}{ds} \cos(\omega^i_l (t-s))dt + F^i_l(t), \\
&F^i_l(t) = \left(x^i_l(0) - \frac{\gamma^i_l}{m {\omega^i_l}^2}X_l(0)\right)\cos(\omega^i_l t) + \frac{\dot{x}^i_l(0)}{\omega^i_l}\sin(\omega^i_l t).
\end{align}

Substituting this solution for $x$ eliminates the fast variable $x$ and yields the following equation 

\begin{align}
  &\dot{X}_l =\color{black}-\color{black} \{H^0_1,H^0_2, X_l\}_{NB} \\
  &- \sum^{3}_{m,n=1}\epsilon_{mln} \int^t_0 \zeta^l(t-s) \frac{\partial H^0_2(t)}{\partial X_n} \frac{d X_l(s)}{d s} ds - \sum^3_{m,n=1} \epsilon_{mln}  \frac{\partial H^0_2(t)}{\partial X_n} F_l, \\
  &\zeta^l(t) = \sum^N_{i=1} \frac{\gamma^i_l}{m {\omega^i_l}^2} \cos(\omega^i_l t), \\
  &F_l = \sum^N_{i=1} F^i_l,
\end{align}
where $F_l$ is the effect of fluctuations involving only degrees of freedom of the heat bath, and $\zeta$ term represents friction. However, since it is reversible as it is, an approximation is made and 
$F$ is taken to be white noise and $\zeta$ term to be
\begin{align}
  - \sum^3_{m,n=1}\epsilon_{mln} \int^t_0 \zeta^l(t-s) \frac{\partial H^0_2(t)}{\partial X_n} \frac{d X_l(s)}{d s} ds \to - \sum^3_{m,n=1}\epsilon_{mln} \zeta_0\frac{\partial H^0_2(t)}{\partial X_n} \frac{d X_l(t)}{d t}, 
\end{align}
where $\zeta_0$ is a continuous approximation of the state of the heat bath and the distribution is assumed to be a Debye distribution, and $\gamma^i_l = \gamma/\sqrt{3N}$.

As an example, the time evolution equation with fluctuation in the case of BZ reaction is given as 
\begin{equation}
\frac{dX}{dt}=k_{1}AY-k_{2}XY+k_{3}AX-2k_{4}X^{2} +\zeta_0 \frac{d Y}{dt} + \kappa  \xi_2,
\end{equation}

\begin{equation}
\frac{dY}{dt}=-k_{1}AY-k_{2}XY+hk_{5}BZ - \zeta_0 \frac{d X} {dt} - \kappa \xi_1,
\end{equation}

\begin{equation}
\frac{dZ}{dt}=2k_{3}AX-k_{5}BZ 
\end{equation}
where $\kappa$ is the diffusion coefficient introduced during the white noise approximation, $F_l = \kappa \xi_l$.

\begin{figure}[t]
    \begin{tabular}{cc}
      \begin{minipage}[t]{0.45\hsize}
        \centering
        \includegraphics[width=1.0\textwidth,height=1.0\linewidth]{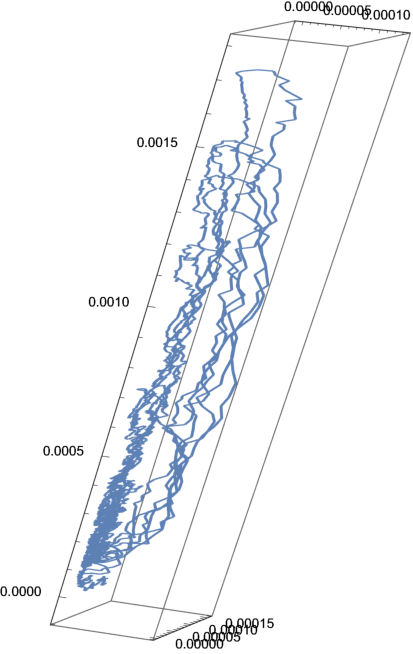}
        \caption{Including fluctuation.}
        \label{fig:case1}
      \end{minipage} &
      \begin{minipage}[t]{0.45\hsize}
        \centering
        \includegraphics[width=1.0\textwidth,height=1.0\linewidth]{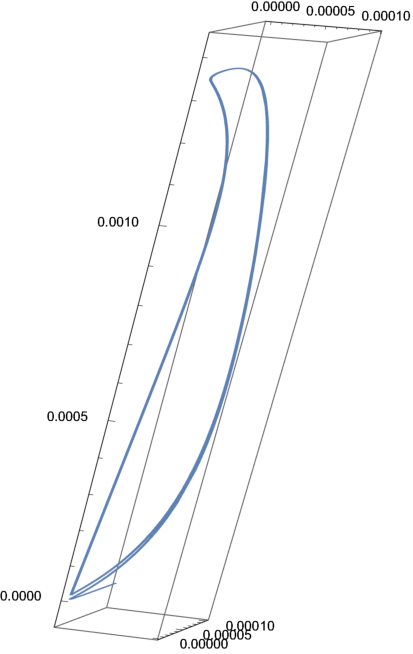}
        \caption{Without fluctuation.}
        \label{fig:Case1.-Non-fluctuation}
      \end{minipage}
    \end{tabular}
  \end{figure}

The numerical results for each case are presented in Figure \ref{fig:case1}.
Figure \ref{fig:Case1.-Non-fluctuation} also shows the path in the
case of no fluctuation for comparison.

The parameters used are as follows: $k_{1}=1.28,\ k_{2}=8.0,\ k_{3}=8.0\times10^5,\ k_{4}=2.0\times 10^3,\ k_{5}=1.0,\ h=1.5,\ A=0.06,\ B=0.02,\ X(0)=2\times 10^{-7},Y(0)=0.0001,\ Z(0)=0.0, \kappa=10^{-7}, \zeta_0 = 0.1$
and $dt=10^{-2}$ for split time interval.

In this case, as can be seen by comparison of the figures, the addition of white
noise simply causes the pathway to fluctuate slightly around the noiseless
case.
\color{black}

\section*{Appendix D. Specific example}
\color{black}
This is a continuation of (Footnote 7).

As a specific example, we can consider the case in which the thermal equilibrium state $\bm{x}_{\text{eq}}$ exists and the system is attracted towards it.  In this case we may choose the following velocity potential, 
\begin{eqnarray}
\phi(x)= - \frac{\lambda_L}{2} (\bm{x}-\bm{x}_{\text{eq}})^2.
\end{eqnarray} 
Then, the velocity field near the equilibrium point $\bm{x}_{\text{eq}}$ becomes $\bm{v} = - \lambda_L (\bm{x}-\bm{x}_{\text{eq}})$, implying the point $\bm{x}_{\text{eq}}$ be a sink or an attractor for $\lambda_L > 0$.  Otherwise it becomes a source or a separation point.  Indeed, the equation of continuity$D_t \; n(t, \bm{x})= \lambda_L N  n(t, \bm{x})$ can be solved, giving the time independent distribution function with a constant $A$,
\begin{eqnarray}
n(t, \bm{x})= f(t, \bm{x})=\frac{A}{ \left[ \left(\bm{x}-\bm{x}_{\text{eq}}\right)^2\right]^{\frac{N}{2}} }.
\end{eqnarray}

Here, we note that the invariant quantity floating along the stream is
\begin{eqnarray}
D_t \; I(t, \bm{x})=0, ~~\text{for}~~ I(t, \bm{x})= e^{\lambda_L t} \sqrt{\left(\bm{x}-\bm{x}_{\text{eq}}\right)^2}.
\end{eqnarray}
This means the distance to the equilibrium point \color{black}behaves \color{black}
\begin{eqnarray}
\sqrt{\left(\bm{x}-\bm{x}_{\text{eq}}\right)^2} \propto  e^{-\lambda_L  t},
\end{eqnarray}
implying we need the infinite time before arriving at the equilibrium.  

We also obtain the behavior like the Lyapunov index for the distribution function,
\begin{eqnarray}
n(t, \bm{x})= f(t, \bm{x}) \propto \frac{1}{ \left[ \left(\bm{x}-\bm{x}_{\text{eq}}\right)^2\right]^{\frac{N}{2}} } \propto e^{N \lambda_L t}.
\end{eqnarray}

Accordingly, $\phi$ is identified to the entropy $S$, since they attain a maximum at the equilibrium point $\bm{x}_{\text{eq}}$.  So, we may call $\lambda_L$ and $\phi(x)$ as Lyapunov exponent and its function.  
 
To understand our choice of velocity potential $\phi$ as entropy $S$, in the standard thermodynamics near equilibrium as an example, the best way is to compare our theory with the Einstein's paper on Brownian motion \cite{Einstein}.  It is because Einstein's paper contains all the necessary ingredients of the present day linear response theory near equilibrium.  He started with the motion of a macroscopic molecule in the solvent.  The Navier-Stokes equation for molecules receive a thermodynamic force $\bm{X}$ from the solvent,  which is balanced with the friction force to the molecular motion $\bm{v}$, with a friction coefficient $\gamma$, as 
\begin{eqnarray}
\gamma \bm{v}= \bm{X} = \frac{\partial S}{\partial \bm{x}},
\end{eqnarray}
where $S$ is the entropy.  This equation is identical to our choice of velocity potential as the entropy, $\bm{v}^{(2)}= \bm{\nabla} \phi = \bm{\nabla} S$.  This is also related to the linear response theory at present, 
\begin{eqnarray}
\bm{j} =n \bm{v}= L  \frac{\partial S}{\partial \bm{x}},
\end{eqnarray}

Einstein considered the non-uniform $n$, so that the more careful examination will reveal the relationship between the random walk or the diffusion process, and the process of approaching to the thermal equilibrium.  In this respect, the difference between $\langle \bm{x}^2 \rangle \propto Nt$ in the random walk and 
$\left(\bm{x}-\bm{x}_{\text{eq}}\right)^2 \propto  e^{-2\lambda_L  t}$ in the process of attarcting to the equilibrium, becomes important.
\color{black}

\color{black}
\section*{Appendix E. Harmonic oscillator}

In the end of Subsec. 5.1, we mentioned that with a special choice of parameters, the B-Z reaction can be described by the harmonic oscillators.  We examine this simplified model in this Appendix, which is described by the following Hamiltonians and entropy:
\begin{align}
    H_1 &= a Y Z, \\
    H_2 &= Z, \\
    S &= \frac{1}{2} b X^2 + c X Z.
\end{align}

The evolution of the system follows the Nambu non-equilibrium thermodynamic equation:
\begin{align}
    \dot{X}^i &= - \{ X^i, H_1, H_2 \} + \frac{\partial S}{\partial X^i}.
\end{align}

For the variables \(X\), \(Y\), and \(Z\), this gives the following equations of motion:
\begin{align}
    \dot{X} &= - a Z + b X + c Z = (c - a) Z + b X, \\
    \dot{Y} &= 0, \\
    \dot{Z} &= c X. \label{H.O.}
\end{align}

We can express this system in matrix form as:
\begin{align}
    \begin{pmatrix}
        \dot{X} \\
        \dot{Z}
    \end{pmatrix}
    &= 
    \begin{pmatrix}
        b & c - a \\
        c & 0
    \end{pmatrix}
    \begin{pmatrix}
        X \\
        Z
    \end{pmatrix}.
\end{align}

Differentiating once more gives:
\begin{align}
    \begin{pmatrix}
        \ddot{X} \\
        \ddot{Z}
    \end{pmatrix}
    &= 
    \begin{pmatrix}
        b & c - a \\
        c & 0
    \end{pmatrix}^2
    \begin{pmatrix}
        X \\
        Z
    \end{pmatrix}
    &=
    \begin{pmatrix}
        b^2 + c(c - a) & b(c - a) \\
        bc & c(c - a)
    \end{pmatrix}
    \begin{pmatrix}
        X \\
        Z
    \end{pmatrix}.
\end{align}

The eigenvalue equation of this matrix:
\begin{align}
    \lambda^2 - (b^2 + 2c(c - a))\lambda - b^2c(c - a) = 0,
\end{align}
gives two real eigenvalues, $\lambda_+$ and $\lambda_-$, if $\lambda_+\lambda_-=-b^2c(c-a) <0$.  Under this condition $c(c-a) >0$, one of the eigen-values $\lambda_{-}$ is negative, whose eigen-mode gives an elliptic trajectory.  That is, there exists a conserved quantity $\tilde{H}_1$, given by a second order algebraic equation for $(X, Z)$.  

Since we know $Y$ is another conserved quantity, which may be denoted by $\tilde{H_2}$, we have arrived at the system having two conserved quantities $(\tilde{H_1}, \tilde{H_2})$.  In this new choice of conserved quantities, no dissipation appears, which implies $\tilde{S}=0$, and the system is described by a pure Nambu dynamics. 

This simplified model shows that even with the entropy $S$, there is a possibility that by changing the choice of Hamiltonians and entropy, we can elucidate the existence of a conservation law in the non-equilibrium thermodynamics with dissipation. This phenomenon was called ''quasi-periodicity'' or emergence of ``quasi-cycle'' in this paper.   It is interesting to note that such phenomenon appearing far from the equilibrium, can be analyzed in the Nambu non-equilibrium thermodynamics.






\color{black}

\end{document}